\documentclass[twocolumn,english,showpacs,preprintnumbers,amsmath,amssymb,floatfix,nofootinbib]{revtex4-1}

\usepackage[T1]{fontenc}
\usepackage[latin9]{inputenc}
\usepackage{color}
\usepackage{array}
\usepackage{amstext}
\usepackage{graphicx}
\usepackage{esint}
\usepackage{rotating}
\usepackage{slashed}
\usepackage{siunitx}
\usepackage[font=small,labelfont=bf]{caption}
\usepackage{xcolor}
\usepackage[colorlinks = true,
linkcolor = magenta,
urlcolor  = blue,
citecolor = red,
anchorcolor = blue]{hyperref}

\usepackage{feynmp}

\makeatletter

\@ifundefined{textcolor}{}
{%
 \definecolor{BLACK}{gray}{0}
 \definecolor{WHITE}{gray}{1}
 \definecolor{RED}{rgb}{1,0,0}
 \definecolor{GREEN}{rgb}{0,1,0}
 \definecolor{BLUE}{rgb}{0,0,1}
 \definecolor{CYAN}{cmyk}{1,0,0,0}
 \definecolor{MAGENTA}{cmyk}{0,1,0,0}
 \definecolor{YELLOW}{cmyk}{0,0,1,0}
 }

\def\missE{\slashed E} 

\@ifundefined{definecolor}
 {\usepackage{color}}{}
\@ifundefined{definecolor}
 {\usepackage{color}}{}
\makeatother
\usepackage{babel}

\newcommand{\delphes}{{\sc Delphes 3.4.2 }}
\newcommand{\madgraph}{{\sc MadGraph5\_aMC@NLO }}

\newcommand{\feynrules}{{\sc Feyn\-Rules }}
\newcommand{\pythia}{{\sc Pythia 8.20 }}
\newcommand{\fastjet}{{\sc FastJet 3.2 }}

\def\Re{{\cal R \mskip-4mu \lower.1ex \hbox{\it e}\,}}
\def\Im{{\cal I \mskip-5mu \lower.1ex \hbox{\it m}\,}}

\def\tev{\,{\ifmmode\mathrm {TeV}\else TeV\fi}}
\def\gev{\,{\ifmmode\mathrm {GeV}\else GeV\fi}}
\def\mev{\,{\ifmmode\mathrm {MeV}\else MeV\fi}}
\def\to{\rightarrow}

\begin{document}


\title { Probing top quark FCNC couplings in the triple-top signal at the high energy LHC and future circular collider }

\author{Hamzeh Khanpour}
\email{Hamzeh.Khanpour@cern.ch}

\affiliation {
Department of Physics, University of Science and Technology of Mazandaran, P.O.Box 48518-78195, Behshahr, Iran        \\
School of Particles and Accelerators, Institute for Research in Fundamental Sciences (IPM), P.O.Box 19395-5531, Tehran, Iran }

\date{\today}

%
\begin{abstract}\label{abstract}

Our main aim in this paper is to present detailed studies to probe the top quark flavor changing neutral current (FCNC) interactions at $tqg$, $tq\gamma$, $tqH$ and $tqZ (\sigma^{\mu \nu}, \gamma_{\mu})$ vertices in the triple-top signal $p p \to t \bar t t \, (\bar t t \bar t)$ at the high energy proposal of Large Hadron Collider (HE-LHC) and future circular hadron-hadron collider (FCC-hh).
To this end, we investigate the production of three top quarks which arises from the FCNC couplings taken into account the fast simulation at $\sqrt{s} = 27$ TeV of HE-LHC and 100 TeV of FCC-hh considering the integrated luminosities of 10, 15 and 20 ab$^{-1}$. All the relevant backgrounds are considered in a cut based analysis to obtain the limits on the anomalous couplings and the corresponding branching ratios. The obtained exclusion limits on the coupling strengths and the branching ratios are summarized and compared in detail with the results in the literature, namely the most recent direct LHC experimental limits and HL-LHC projections as well. We show that, for the higher energy phase of LHC, a dedicated search for the top quark FCNC couplings can achieve much better sensitivities to the triple-top signal than other top quark production scenarios. We found that the limits for the branching ratios of $tqg$ and $tqH$ transitions could reach an impressive sensitivity and the obtained 95\% CL limits are at least three orders of magnitude better than the current LHC experimental results as well as the existing projections of HL-LHC.

\end{abstract}
%


\maketitle
\tableofcontents{}

%
\section{ Introduction and motivations }\label{sec:intro}

The top quark with the mass of $m_{\rm top} = 173.0 \pm 0.4 \, {\text {GeV}}$~\cite{Tanabashi:2018oca} which is close to the electroweak symmetry breaking scale, is the most sensitive probe to search for a new physics evidence beyond the Standard Model (SM) at hadron and lepton colliders~\cite{Husemann:2017eka,Giammanco:2017xyn}. Due to the large mass of the top quark, the productions and related theoretical and experimental studies are golden places to look for possible signatures of new physics at TeV scales. Whilst improving the precision of SM predictions is highly important in its own right, any studies on the top quark to probe signatures of new physics are also most welcome. In this respect, over the past few years, several dedicated studies have shown that the non-SM couplings of the top quark should be one of the key analysis program pursued at the Large Hadron Collider (LHC)~\cite{Husemann:2017eka,Englert:2017dev,Cortiana:2015rca,Boos:2015bta,Cristinziani:2016vif}. These dedicated studies have been done in the top quark related processes, most notably in the single top quark production~\cite{Giammanco:2015bxk} or in the double top quarks production~\cite{Gouz:1998rk,Wicke:2010xw} scenarios. Among them, the flavor changing neutral current (FCNC) interactions involving a top quark, other quark flavors and neutral gauge boson are much of interest.

The FCNC interactions of top quark are forbidden at the tree level and due to the Glashow-Iliopoulos-Maiani mechanism (GIM)~\cite{Glashow:1970gm}, are highly suppressed in a loop level. The SM predictions of the top quark FCNC decays to the gluon, photon, Z and Higgs boson, and an up or charm quark are expected to be at the order of ${\cal O} (10^{-12}-10^{-17})$ which are currently out of range of present and even future experimental sensitivity~\cite{AguilarSaavedra:2004wm}.
However, in the beyond SM scenarios such suppression can be relaxed which yield to couplings of the orders of magnitude larger than those of the SM~\cite{Gaitan:2017tka,Chiang:2018oyd,AguilarSaavedra:2002ns,Cao:2007dk,Baum:2008qm,Agashe:2004cp,Cao:2007bx,Zhang:2007ub}. Hence, the possible deviation from the SM predictions of FCNC couplings would imply the existence of new physics beyond the SM~\cite{Oyulmaz:2018irs}. In recent years, 
there has been a growing number of analyses focusing on this topic, and to date, there are many phenomenological studies in literature that have extensively investigated the association production of top quark with a gluon, photon, Higgs and $Z$ boson mainly through single or double top production at hadron and lepton colliders, see e.g. Refs.~\cite{Oyulmaz:2018irs,Khatibi:2014via,Khatibi:2014bsa,Khatibi:2015aal,TurkCakir:2017rvu,Denizli:2017cfx,Khanpour:2014xla,Hesari:2015oya,Shi:2019epw,Alici:2019asv,Oyulmaz:2019jqr,Behera:2018ryv,Aguilar-Saavedra:2017vka,Shen:2018mlj,AguilarSaavedra:2010rx,Jain:2019ebq,Liu:2020kxt} for the most recent reviews.

At the LHC, the top pair $p p \to t \bar t$ and single-top quark productions are the dominated processes due to the strong coupling of $g g \to t \bar t$ subprocess~\cite{Giammanco:2015bxk,CMS:1900mtx}. The production of an odd number of top quarks, i.e. triple-top quarks $p p \to  t \bar t  t \, (\bar t t \bar t)$, requires a $t b W$ vertex in every diagram. Since it also often involves a $b$-quark in the initial state of the hard process, therefore, they lead to a significant suppression in comparison to the strong processes. At the 14 TeV energy of LHC, triple-top quarks production cross section, with $\sigma \simeq 1.9 \, {\rm fb}$, is almost five orders of magnitude less than the top pair production which is the dominant mechanism of top productions at the LHC. This relatively small SM production cross section of three top quarks makes it an interesting channel for investigating any signal of new physics~\cite{Cao:2019qrb}. 

The LHC at CERN and its luminosity upgrade, (HL-LHC)~\cite{CidVidal:2018eel,Cepeda:2019klc,Cepeda:2019klc,Azzi:2019yne,CMS:2018kwi,Cerri:2018ypt,ATLAS:HL-LHC}, have been actively carried out and will still continue the journey on searching for any signal of new physics in the next two decades.
In addition to the HL-LHC, there are other proposals for the future higher energy hadron colliders to perform the direct searches at the energy frontier. These proposals include the energy upgrade for the LHC to 27 TeV center-of-mass energy (HE-LHC)~\cite{CidVidal:2018eel,Cepeda:2019klc,Azzi:2019yne} and the future circular collider of about 100 TeV center-of-mass energy FCC-hh~\cite{Benedikt:2018csr,Mangano:2017tke,Arkani-Hamed:2015vfh}. They will collect datasets corresponding to integrated luminosities of 10-20 ab$^{-1}$ and 10 ab$^{-1}$, respectively. The high energy and high luminosity reach of these colliders could open a large window for new physics research and strongly motivated to search for any possible signature of extremely rare top quark FCNC couplings. The FCC-hh is considered to be able to discover certain rare processes, new interactions up to masses of $\sim$30 TeV and search for a new physics. Due to the great center-of-mass energy and collision rate, large amounts of top quark and Higgs boson will be produced, which provide an unbeatable opportunity to search for the $t \to qX$ FCNC couplings.  Considering the need of these proposed colliders and our discussion on three top productions, one can conclude that the triple-top signal at HE-LHC or FCC-hh may potentially provide clear evidence for the top quark FCNC couplings. In this paper, we set out an initial study for the triple-top signal and present a detailed study to probe top quark FCNC interactions at $t q g$, $t q \gamma$, $t q H$ and $t q Z (\sigma^{\mu \nu}, \gamma_{\mu})$ transitions at HE-LHC and FCC-hh. It should be mentioned here that, the triple-top quark signal that we are interested to investigate makes it possible to study all the top quark FCNC interactions $tqX$. 

On the experimental side, lots of efforts performed earlier at the Tevatron at Fermilab and now at the 13 TeV LHC have failed to reveal any interesting observation
of FCNC transitions. However, the obtained bounds on such couplings from the mentioned experiments are very strong. Most recently, considering the 13 TeV data from CMS and ATLAS, the exclusion limits on the top quark FCNC transitions have significantly improved by the LHC experiments. 
CMS and ATLAS Collaborations at CERN reported the most stringent constrains through the direct measurements~\cite{Aad:2015gea,Khachatryan:2015att,Sirunyan:2017kkr,Aaboud:2017mfd,Khachatryan:2016sib,Aaboud:2018pob,CMS:2017twu,CMS:2016bss,Aaboud:2018nyl,Sirunyan:2017nbr,Aaboud:2017ylb,Aaboud:2018oqm,Sirunyan:2017uae,Aad:2015pja}.  

These collaborations have set upper limits on the $tqH$ FCNC couplings in the top sector at $\sqrt{s} = 13 \, {\rm TeV}$ considering an integrated luminosity of 36.1 fb$^{-1}$ (ATLAS) and 35.9 fb$^{-1}$ (CMS).  Considering the analyses of the different top FCNC decay channels, the 95\% confidence  level (CL) upper limits have been found to be  $Br (t \to u H) < 0.19\%$ and  $Br (t \to c H) < 0.16\%$ from the ATLAS~\cite{Aaboud:2018pob}, and $Br (t \to u H) < 0.34\%$ and  $Br (t \to c H) < 0.44\%$ from the CMS~\cite{Sirunyan:2017uae} Collaborations. In addition to this direct collider measurement for $tqH$ couplings, single top quark production in the $t$ channel is used to set limits for the top quark FCNC interactions with gluon $tqg$ considering the data taken with the CMS detector at 7 and 8 TeV correspond to the integrated luminosities of 5.0 and 19.7 fb$^{-1}$. The upper limits on the branching fractions of  $Br (t \to u g) < 0.002\%$ and  $Br (t \to c g) < 0.041\%$ have been measured~\cite{Khachatryan:2016sib}. A search for FCNC through single top quark production in association with a photon also have been performed by CMS at $\sqrt{s} = 8 \, {\rm TeV}$ corresponding to an integrated luminosity of 19.8 fb$^{-1}$. Upper limits at the 95\% CL on $tq\gamma$ anomalous couplings are measured to be $Br (t \to u \gamma) < 0.013\%$ and  $Br (t \to c \gamma) < 0.17\%$~\cite{Khachatryan:2015att}. Finally, search for the FCNC top quark decays of $t \to qZ$ in proton-proton collisions at $\sqrt{s} = 13 \, {\rm TeV}$ have been done both by CMS and ATLAS Collaborations through different channels. Upper limits at 95\% CL level on the branching fractions of top quark decays can be found to be $Br (t \to u Z) < 0.015\%$ and  $Br (t \to c Z) < 0.037\%$ from the CMS~\cite{CMS:2017twu}, and $Br (t \to u H) < 0.024\%$ and  $Br (t \to c H) < 0.032\%$ from the ATLAS~\cite{Aaboud:2018nyl} Collaborations for the integrated luminosities of 35.9 and 36.1 fb$^{-1}$, respectively.

In this paper, we shall try to investigate in details the projected sensitivity and discovery prospects of the HE-LHC and FCC-hh to the top quark FCNC transitions within the model independent way using an effective Lagrangian framework. To this end, we follow the strategy presented in~\cite{AguilarSaavedra:2008zc,AguilarSaavedra:2004wm} and quantify the expected sensitivity of the HE-LHC and FCC-hh to the top quark FCNC couplings $tqX$. The realistic detector effects are included in the production of the signal and background processes with the most up-to-date experimental studies carefully considering the upgraded CMS detector performance~\cite{Atlas:2019qfx} for the HE-LHC and the FCC-hh baseline detector configuration embedded into Delphes. As we will demonstrate, the expected constraints for $tqg$ and $tqH$ from the HE-LHC and FCC-hh are significant and fully complementary with those from the LHC and HL-LHC. We show that the $tqg$ and $tqH$ couplings can be constrained to the order of ${\cal O} (10^{-6})$\% and ${\cal O} (10^{-3})$\%. These estimates are two to three 
order of magnitude more stringent than LHC and HL-LHC extrapolations. Our findings indicate that the triple-top quark production could provide 
complementary information to the new physics searches in the same-sign top quark pairs.

This article is arranged as follows. In Sec.~\ref{sec:framework}, we present the theoretical framework and the effective Lagrangian approach for the top quark FCNC couplings. The details of the analysis strategy applied in this investigation are clearly discussed and presented in Sec.~\ref{sec:strategy}. This section also includes the signal and background estimations, the simulations and detector effects for HE-LHC and FCC-hh.
We detail in Sec.~\ref{sec:statistical} the statistical method we assume, together with the numerical calculations and distributions for the HE-LHC and FCC-hh. Sec.~\ref{sec:results} includes the numerical results and findings in details. The 95\% confidence level (CL) limits of HE-LHC and FCC-hh are compared with the LHC measured limits and the other studies in literature. Finally, in Sec.~\ref{sec:Discussion}, we conclude and summarize our main results and findings.


%
\section{Theoretical framework and assumptions}\label{sec:framework}
%

This section presents the theoretical framework and assumptions in which we rely for our research to study the top quark FCNC transitions at HE-LHC and FCC-hh.
The possibility of the top quark anomalous FCNC couplings with light quarks $(q=u, c)$ and a 
gauge bosons ($g, \, H, \, Z, \, \gamma$) is explored in a model-independent way considering the most general effective Lagrangian approach~\cite{AguilarSaavedra:2008zc,AguilarSaavedra:2009mx}. In the search of anomalous FCNC interactions at high energy colliders, this approach 
has been extensively studied in literature for lepton and hadron colliders~\cite{Oyulmaz:2018irs,Khatibi:2014via,Khatibi:2014bsa,Khatibi:2015aal,TurkCakir:2017rvu,Denizli:2017cfx,Khanpour:2014xla,Hesari:2015oya,Shi:2019epw,Alici:2019asv,Oyulmaz:2019jqr,Behera:2018ryv,Goldouzian:2014nha,Goldouzian:2016mrt,Buschmann:2016uzg,Khatibi:2015aal,delAguila:1999kfp,Sun:2016kek,Guo:2016kea,Liu:2016gsi,Liu:2016dag,Liu:2019wmi,Shi:2019epw,Kumbhakar:2019njm,Malekhosseini:2018fgp,Papaefstathiou:2017xuv,Dey:2016cve}. In this framework, these FCNC vertices are described by higher-dimensional effective operators ${\cal L}_{\rm FCNC}^{tqX}$ independently from the underlying theory. Up to dimension-six operators, the FCNC Lagrangian of the $tqg$, $tqH$, $t q Z (\sigma^{\mu \nu})$, $t q Z (\gamma_{\mu})$ and $tq\gamma$ interactions can be written as~\cite{AguilarSaavedra:2004wm,AguilarSaavedra:2008zc,AguilarSaavedra:2009mx}:

\begin{equation}
\label{Effective-Lagrangian}
\begin{split}
\mathcal{L}_{\text 
{FCNC}}^{tqX} = &
\ \sum_{q = u, c}
\bigg[
\frac{g_s}{2 m_{t}}
\bar{q} \lambda^{a}
\sigma^{\mu \nu} \,
 (\zeta_{qt}^{L} P^{L}
+  \zeta_{qt}^{R} P^{R}) \, 
t G^{a}_{\mu \nu} \\
&\ - \frac{1}{\sqrt{2}}
\bar{q} \, (\eta_{qt}^{L} P^{L}
+ \eta_{qt}^{R} P^{R}) \, t H   \\
&\ + \frac{g_W}{4 c_{W} m_{Z}}
\bar{q} \sigma^{\mu \nu} \,
(\kappa_{qt}^{L} P_{L}
+  \kappa_{qt}^{R} P_{R})
\, t Z_{\mu \nu}   \\
&\ - \frac{g_W}{2 c_{W}} \bar{q}
\gamma^{\mu} \, (X_{qt}^{L} P_{L} +
X_{qt}^{R} P_{R}) \, t Z_{\mu}   \\
&\ + \frac{e}{2 m_{t}} \bar{q}
\sigma^{\mu \nu} \,
  (\lambda_{qt}^{L} P_{L}
+  \lambda_{qt}^{R} P_{R}) \,
t A_{\mu \nu}  \bigg]
+ \text{h.c.} \,.
\end{split}
\end{equation}

In Eq.~\eqref{Effective-Lagrangian}, the real parameters $\zeta_{qt}$, $\eta_{qt}$, $\kappa_{qt}$, $X_{qt}$ and $\lambda_{qt}$ represent the strength of FCNC interactions of a top quark with gluon, Higgs, Z and $\gamma$, respectively. $q$ indicates an up or charm quark as well. 
At the tree-level, in the SM, all the above coefficients are zero and in the presence of the anomalous FCNC vertices the straightforward way is to set limits on these couplings strength and the corresponding branching fractions. In the above equation, $g_s$ is the strong coupling constant and $P_{L(R)}$ denotes the left (right) handed projection operators. In this study, we assume no specific chirality for the anomalous FCNC interactions, and hence, we set $\zeta_{qt}^{L} =  \zeta_{qt}^{R} = \zeta_{qt}$,  $\eta_{qt}^{L} =  \eta_{qt}^{R} = \eta_{qt}$,  $\kappa_{qt}^{L} =  \kappa_{qt}^{R} = \kappa_{qt}$,  $X_{qt}^{L} =  X_{qt}^{R} = X_{qt}$  and $\lambda_{qt}^{L} =  \lambda_{qt}^{R} = \lambda_{qt}$. As we mentioned earlier in the introduction, the triple-top quark signal includes all the top quark FCNC interactions, and hence, make it possible to study all these coefficients in this signal topology.

\begin{figure*}[htb]
\begin{center}
\vspace{0.50cm}
\resizebox{0.40\textwidth}{!}{\includegraphics{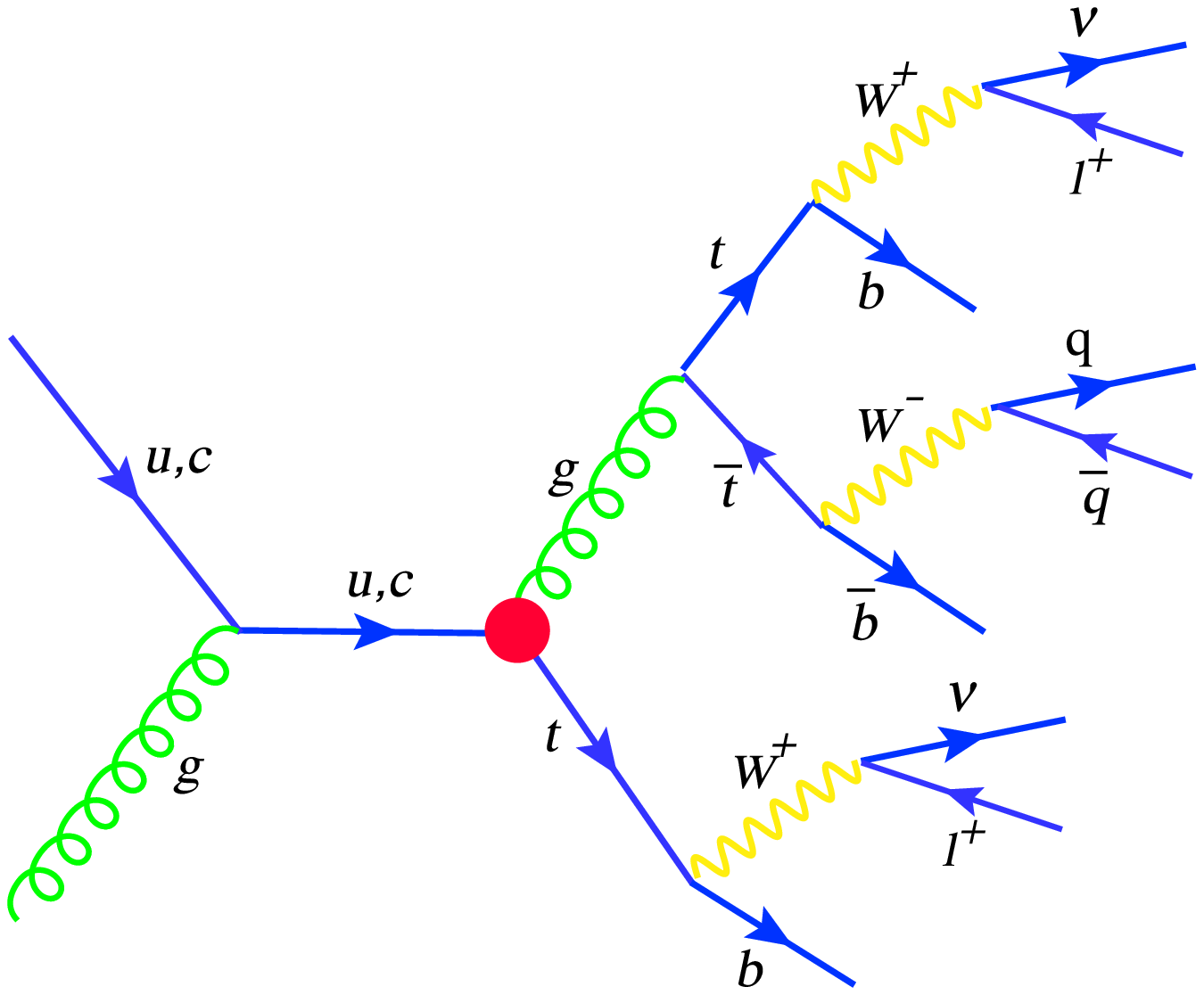}}   		
\resizebox{0.40\textwidth}{!}{\includegraphics{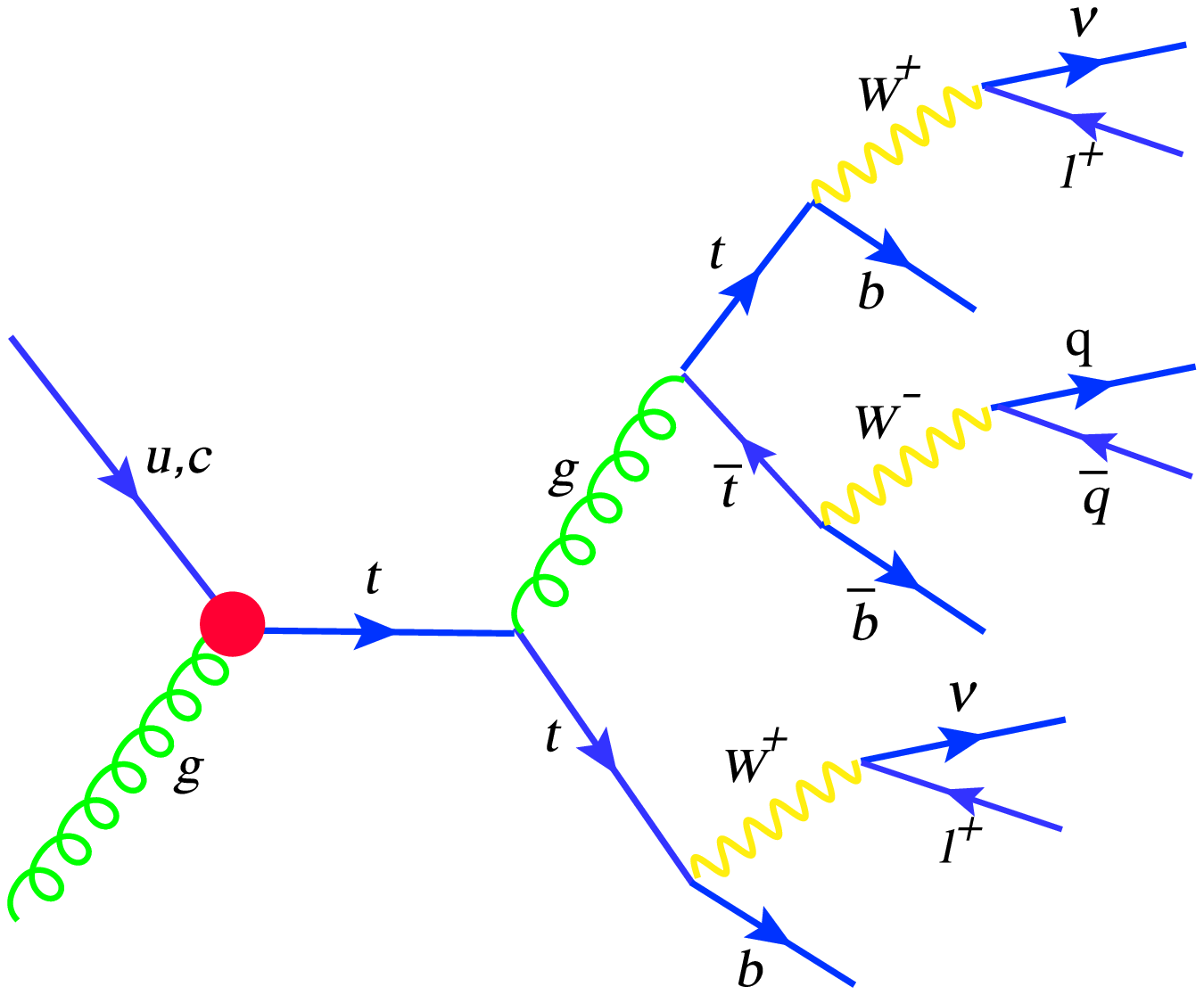}} 
\caption{ The Feynman diagram for the triple-top quark production containing $tqg$ anomalous FCNC vertex. As we described in the text, we consider the leptonic decays of the W boson originating from the same-sign top quarks which lead to same-sign dilepton final states.   } \label{fig:Feynman-Sginal-1}
\end{center}
\end{figure*}

\begin{figure*}[htb]
\begin{center}
\vspace{0.50cm}
\resizebox{0.30\textwidth}{!}{\includegraphics{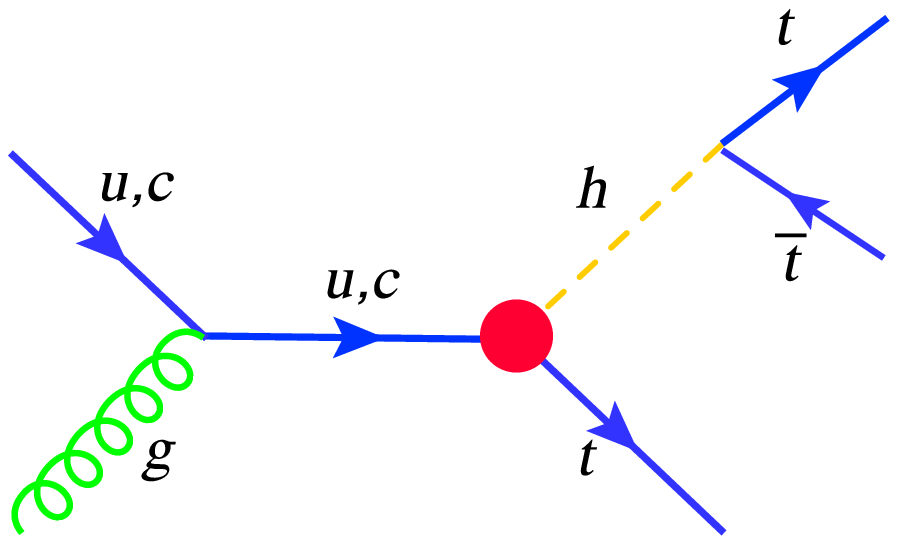}}   	
\resizebox{0.30\textwidth}{!}{\includegraphics{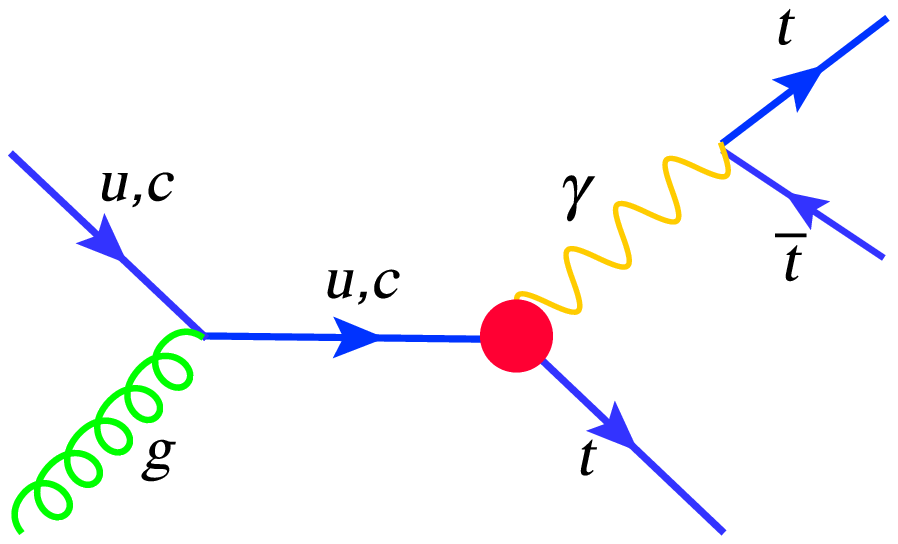}}   
\resizebox{0.30\textwidth}{!}{\includegraphics{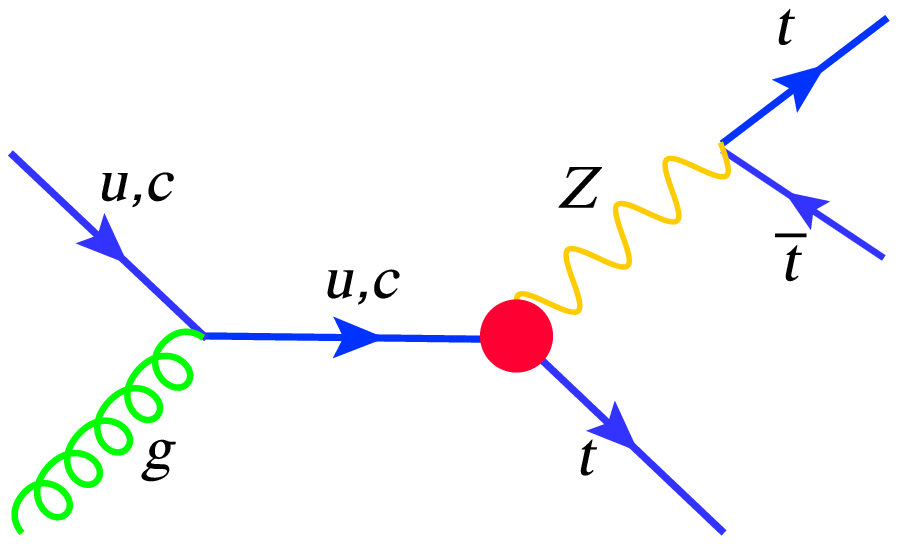}}   
\caption{ The Feynman diagrams for the triple-top quark production in the presence of FCNC $q \to t H$, $q \to t \gamma$ and $q \to t Z$ vertices.  } \label{fig:Feynman-Sginal-2}
\end{center}
\end{figure*}

In this study, we consider $p p \to t \bar t t \, (\bar t t \bar t)$ signal process to search for anomalous FCNC $tqX (X = g, H, Z, \gamma)$ interactions in the presence of effective Lagrangian of Eq.~\eqref{Effective-Lagrangian}. In order to provide more details on the FCNC vertices, we present in Fig.~\ref{fig:Feynman-Sginal-1}, representative Feynman diagrams contributing to this signal process at tree level. As one can see from Fig.~\ref{fig:Feynman-Sginal-1}, this Feynman diagram contains $t q g$ vertices (red circle) in which make it possible to study the top quark FCNC coupling in this signal process. We consider the leptonic decays of the W boson originating from the same-sign top quarks which lead to same-sign dilepton final states. Another top quark decays hadronically. In order to provide more insight on the available FCNC transitions in triple-top productions, we present in Fig.~\ref{fig:Feynman-Sginal-2}, a set of Feynman diagrams contributing to FCNC vertices $q \to t H$, $q \to t \gamma$ and $q \to t Z$. The FCNC vertices are shown as a red circle.

\begin{figure*}[htb]
\begin{center}
\vspace{0.50cm}
\resizebox{0.48\textwidth}{!}{\includegraphics{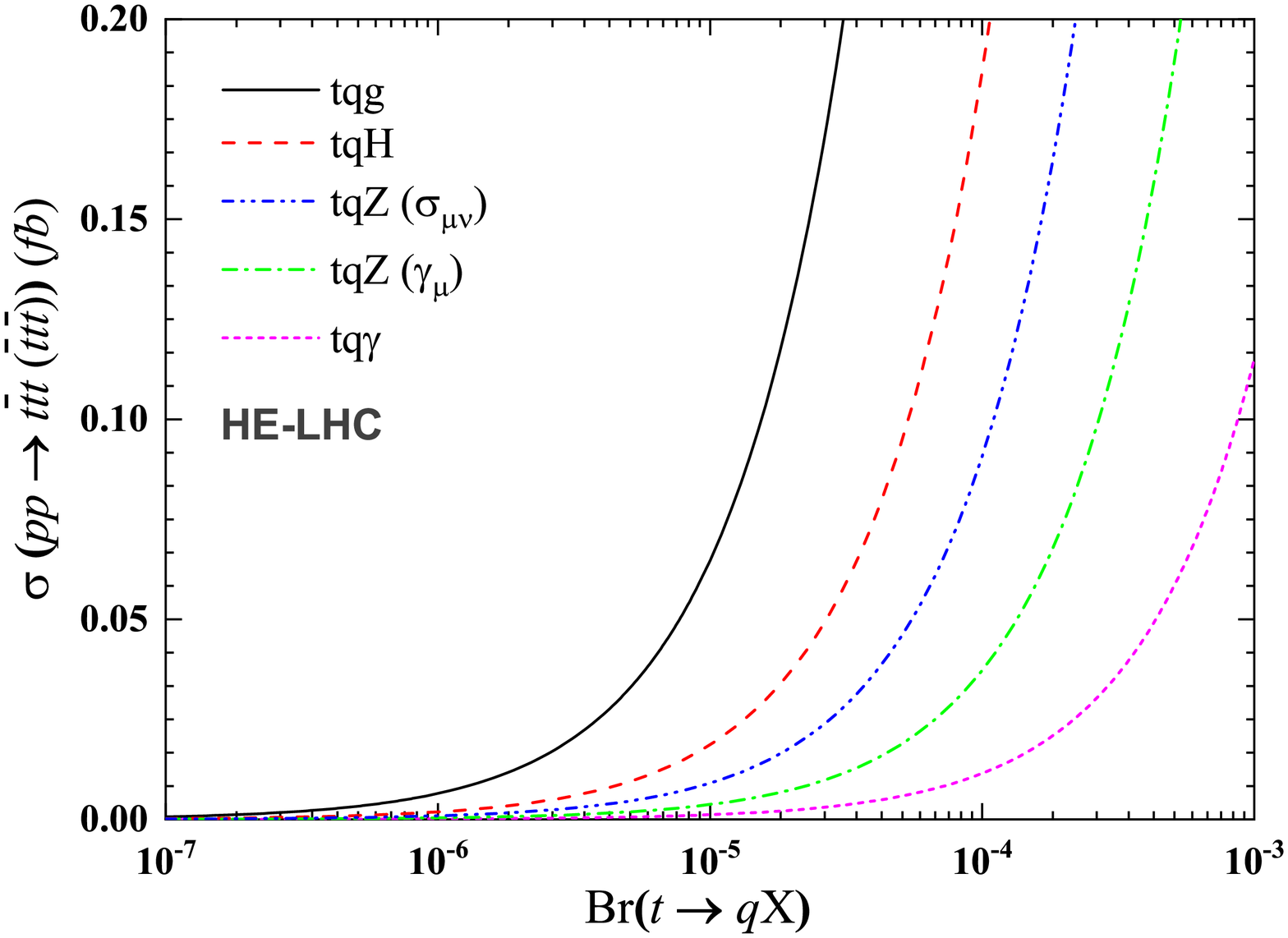}}   	
\resizebox{0.48\textwidth}{!}{\includegraphics{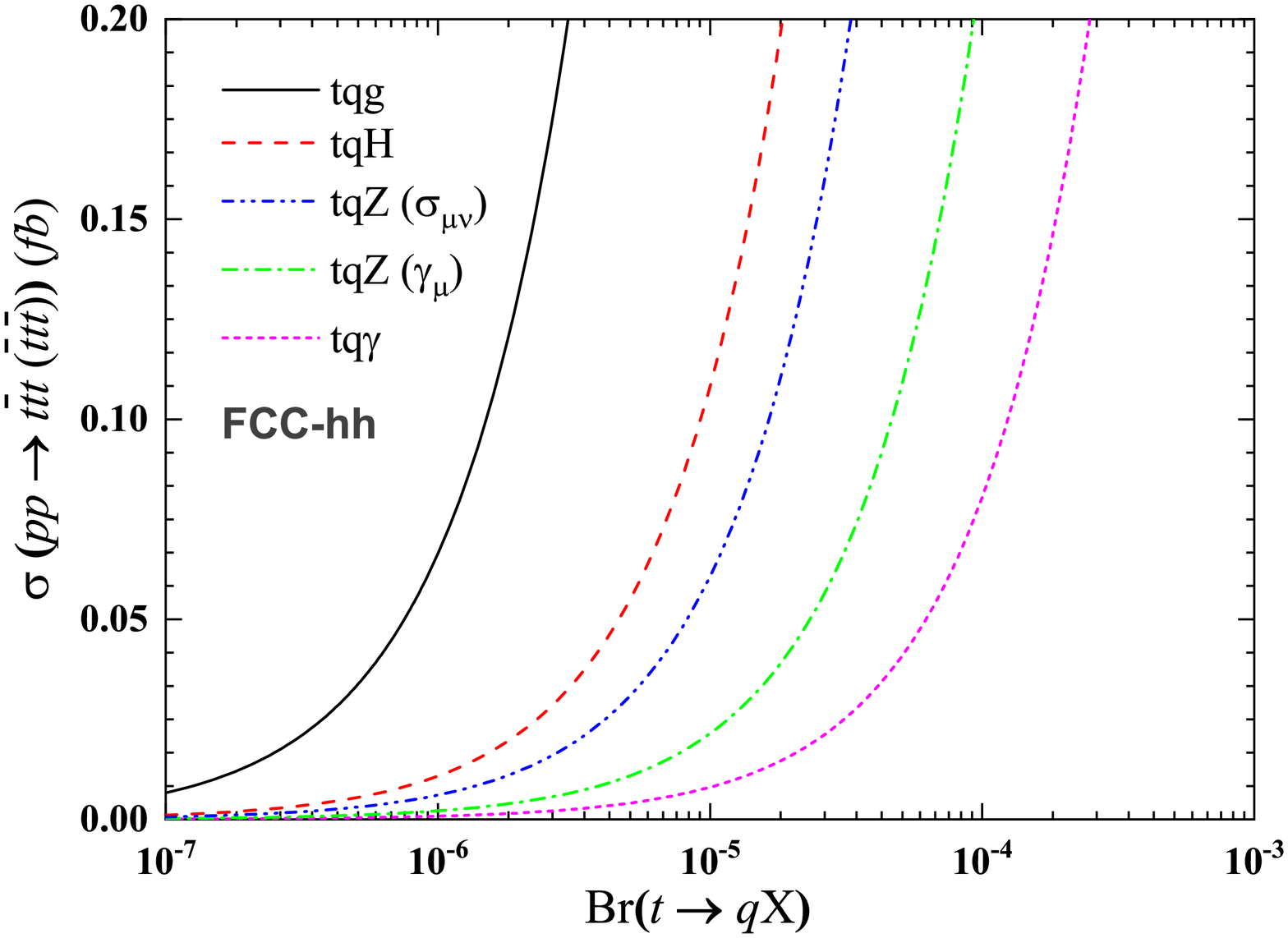}}   	
\caption{ The total cross sections of anomalous $tqX$ couplings $\sigma(pp \to t \bar t t (\bar t t \bar t))$ in the unit of {\it {fb}} for HE-LHC (left) and FCC-hh (right) versus the top quark FCNC branching ratios $Br (t \rightarrow q X)$ for five different signal scenarios. } \label{fig:Sigma_Br}
\end{center}
\end{figure*}

In Fig.~\ref{fig:Sigma_Br}, we show the total cross sections $\sigma(pp \to t \bar t t (\bar t t \bar t))$ in the unit of {\it {fb}} in the presence of anomalous $tqX$ couplings versus the top quark FCNC branching ratios $Br (t \rightarrow q X)$ for five different signal scenarios of $tqg$, $tqH$, $tqZ (\sigma^{\mu \nu}, \gamma_{\mu})$ and $tq\gamma$.

The following conclusions can be drawn from the results presented in Fig.~\ref{fig:Sigma_Br}. As it can be seen, in term of individual $tqX$ coupling, the largest contributions for the triple-top signal mainly come from the $tqg$ coupling, and then the $tqH$. This indicates the large parton distribution functions (PDFs) of $u$-quark and gluon in the calculation of cross section at high center-of-mass energy. In our study, we show that the sensitivity to the branching ratio of $tqg$ and $tqH$ channels are much better than the current LHC experimental limits, and even they are much better than the projected limits on top FCNC couplings at HL-LHC with an integrated luminosity of ${\cal L}_{int} = 3000$ fb$^{-1}$. These findings suggest that the measurement of $tqg$ and $tqH$ FCNC couplings through triple-top productions at future high energy collider would carry a significant amount of information, and hence, are the most sensitive probe to search for a new physics beyond the SM. Fig.~\ref{fig:Sigma_Br} also reflects the less sensitivity of triple-top signal analyzed in this study to the $t q \gamma$ transition.


%
\section{  Analysis strategy and numerical calculations }\label{sec:strategy}
%

As we mentioned, in this study, we plan to investigate the discovery potential of future HE-LHC with 27 TeV C.M. and FCC-hh with 100 TeV C.M. to the top quark FCNC transitions. To this end, we follow an strategy based on an effective Lagrangian approach to describe the top quark FCNC in a model independent way. There are a lot of studies in literature that have been done in new physics searches to enrich the physics motivations of such proposed colliders~\cite{CidVidal:2018eel,Atlas:2019qfx,Cepeda:2019klc,Azzi:2019yne}. One of our main goals in this paper is to study the impact of HE-LHC and FCC-hh to the top quark FCNC coupling determinations. After introducing our theoretical framework and assumptions in previous section, we now present the analysis strategy and numerical calculations related to our study. We first discuss the $tqX$ signal and background analysis. Then, we present the simulation and realistic detector effects for both HE-LHC and FCC-hh.

%
\subsection{ The $tqX$ signal and SM backgrounds  }\label{sec:signal-and-background}
%

In the following section, our study on the $p p \to t \bar t t (\bar t t \bar t)$ signal process including the FCNC $tqX (X = g, H, Z, \gamma)$ couplings as well as the relevant SM backgrounds at the HE-LHC and FCC-hh are given. This signal provides searching for all FCNC couplings of $tqg$, $tqH$, $tqZ$ and $tq\gamma$ independently. We also do this study separately considering $q=u$ and $q=c$. For the triple-top quark, we consider both hadronic ($j j$) and leptonic ($\ell \nu$) decays of $W$ boson by analyzing a very clean signature with two same-sign leptons ($\ell^{\pm \pm }$), where the lepton could be an electron or a muon. Then, the signal analysis is performed with the following final states: $p p \to t \bar t t \to W^+ (\ell^+ \nu_\ell)  b \, \,  W^- (j j) \bar b \, \, W^+ (\ell^+ \nu_\ell) b$ and $p p \to \bar t t \bar t \to W^- (\ell^- \bar \nu_\ell)b \,\, \bar W^+(jj) b \, \, W^- (\ell^- \bar \nu_\ell) b$. Therefore, these unique signal events are characterized by the presence of exactly two isolated same-sign charged leptons, ($2 \ell^+$ or $2 \ell^-$). In addition, there should be a large missing transverse energy (MET) from the undetected neutrino. This signal also characterized by several jets in which three of them should come from $b$-quarks. As one can see from the Feynman diagrams (see Figs.~\ref{fig:Feynman-Sginal-1} and \ref{fig:Feynman-Sginal-2}), the top quark FCNC couplings can be understood by considering the appearance of subprocess diagrams like $tqX \to \bar t t \bar t (t \bar t t)$ with $q=u, c$ and $X = g,\, H,\, Z,\, \gamma$. 

Considering these signal scenarios, the following relevant background processes which have similar final state topology need to be taken into account: 
$t \bar t Z$ in which $Z$ decays to a pair of opposite-sign isolated leptons ($Z \to \ell^+ \ell^-$) with semi-leptonic decay of one top quark and fully hadronic decay of another top quark. We also consider the $ t \bar t W$ with leptonic decay of $W$ boson ($W \to \ell \nu$), semi-leptonic of one top and fully hadronic decay of another quark.
In addition, the  $WWZ$ background also included in which $Z$ decays to a pair of leptons ($Z \to \ell^+ \ell^-$) and the $W$ decays to quark-antiquark pair (hadronic) or to a charged lepton ($\ell^\pm$) and a neutrino (leptonic).

We also consider the SM three top production ($t \bar t t$, $\bar t t \bar t$) as an another important source of background for the signal signature that we considered in this study. 
At leading order (LO) in SM, triple-top signal are produced in association with a $b$-jet  ($t \bar t t + \bar b$, $\bar t t \bar t +b$), or a light-jet ($t \bar t t + \text{light-jet}$, $\bar t t \bar t + \text{light-jet}$), or associated with a W boson ($t \bar t t + W^-$, $\bar t t \bar t + W^+$)~\cite{Barger:2010uw,Cao:2019qrb,Chen:2014ewl}. It has shown that at the $\sqrt{s} = 14$ TeV of LHC energy, the cross section for the triple-top production, with $\sigma_{(t \bar t t, \bar t t \bar t)} = 1.9$ (fb) which is quite negligible, is five orders of magnitude less than the dominant mode of top creation at the LHC, \textit{i.e.} the top-pair production~\cite{Barger:2010uw}. However, the total cross-sections considering the leptonic decay ($\ell \nu_{\ell}$) and hadronic decay ($j j$) of $W$ boson, semileptonic decays of two same-sign top quarks and fully hadronic decay of another top quark are found to be relatively small for the case of HE-LHC. We found that the cross section for the FCC-hh energy is much larger than HE-LHC and need to be taken into account.

Top pair $t \bar t$ events in dilepton channel also can contribute to the backgrounds however the detailed estimation shows that it is negligible and does not have a considerable contribution. We should highlight here that there are contributions from top pair events in dilepton channel when the charge of a lepton is mismeasured. As it is mentioned in Ref.~\cite{Chatrchyan:2012fla}, the probability for charge mismeasurement for muons ($\mu$) is quite small and negligible but it is calculated to be at the order of $(3.3 \pm 0.2) \times 10^{-4}$  for the electrons ($e$). We have carefully investigated this background and found that the requirements of having at least five jets $(n^{\rm jets} \geq 5)$ from which at least three must be $b$-tagged jets $(n^{\rm b-jets}  \geq 3)$, and very small charge mismeasurement probability for electrons significantly suppress the contribution of $t \bar{t}$ events in dilepton channel. It is found that the cross section of  $t \bar{t}$ background after our selection criteria  is very small and can safely ignore in this study. 

Finally, we consider the SM four top production ($t \bar t  t   \bar t$) with both semi-leptonic decay and fully hadronic decay of two same-sign top quarks as an important source of relevant potential SM background~\cite{CMS:1900mtx}. We examined in details the SM four top production background indicating that the contribution from this background need to be taken into account, specially  for the FCC-hh energy. 

\begin{figure*}[htb]
\begin{center}
\vspace{0.50cm}
\resizebox{0.30\textwidth}{!}{\includegraphics{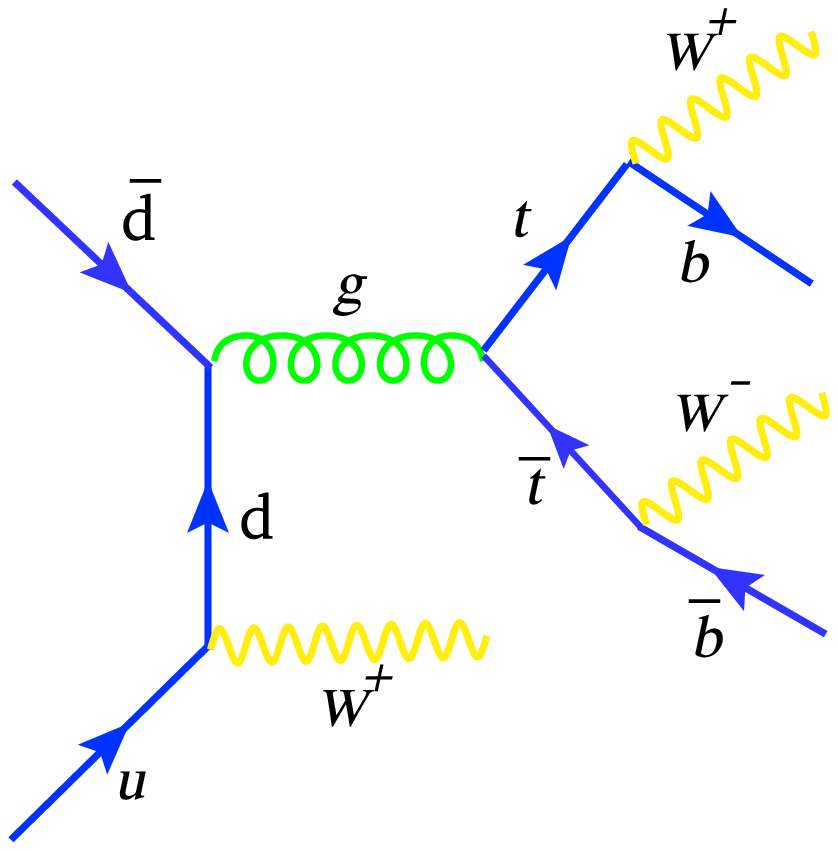}}
\resizebox{0.25\textwidth}{!}{\includegraphics{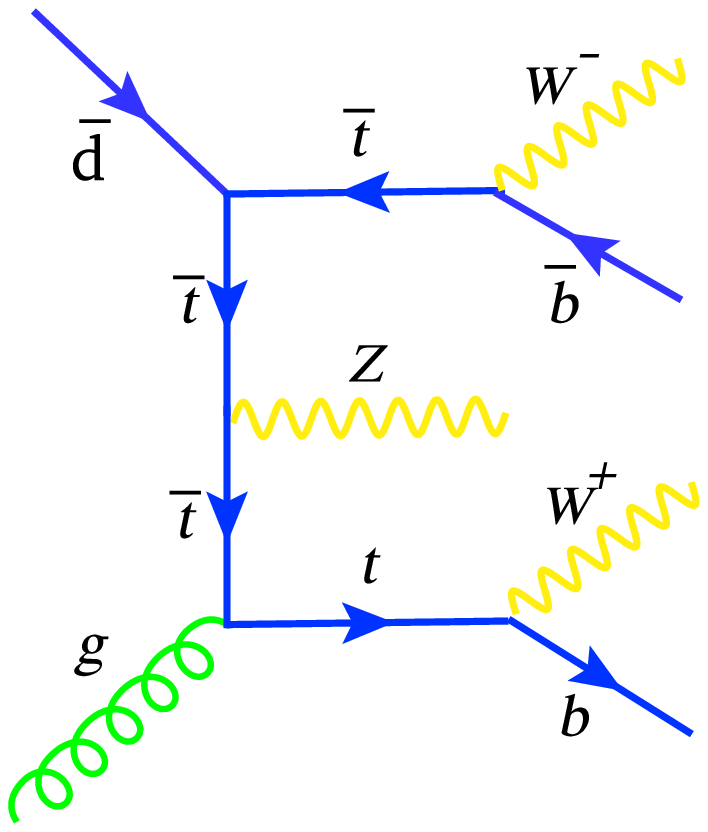}}
\resizebox{0.20\textwidth}{!}{\includegraphics{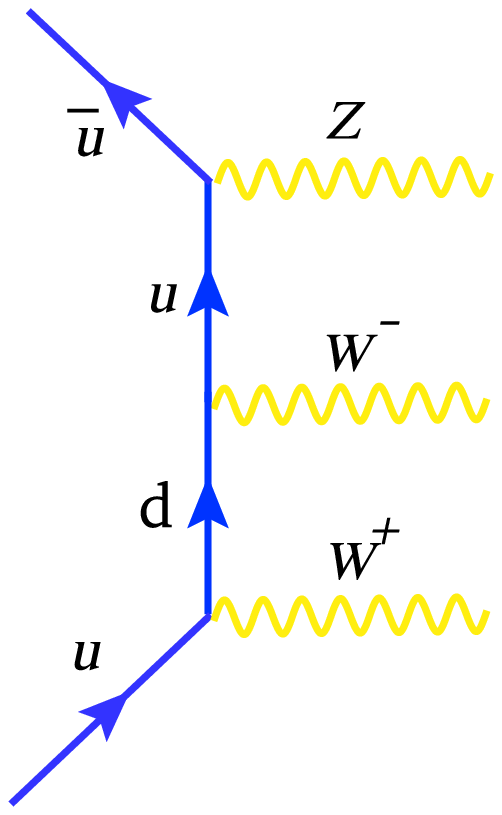}}
\caption{ Some selected examples of partonic Feynman diagrams for the $t \bar t W$, $t \bar t Z$ and $WWZ$ backgrounds considered in this study. The details of background selection are contained in the text.  } \label{fig:Feynman-Back}
\end{center}
\end{figure*}

The backgrounds mentioned above are the most important sources of backgrounds analyzed in this study. In addition to these relevant backgrounds, we also consider other source of backgrounds such as  $t \bar t H$, $t \bar t W W$ and $t \bar t W W$ which are found to be relatively small for the case of HE-LHC and FCC-hh. In order to minimize the contributions of these backgrounds, different selection cuts are applied which will be discussed in details in Sec.~\ref{sec:statistical}. We show that by applying a same-sign isolated dilepton and 3 $b$-jets selections, some of these SM backgrounds can be strongly reduced and safely ignored, and hence, they are not considered. Some selected examples of partonic Feynman diagrams for the  $\bar t t W$, $t \bar t Z$ and $WWZ$ backgrounds analyzed in this study are shown in Fig.~\ref{fig:Feynman-Back}.

In the next section, we present the simulation of signal ans backgrounds, and the realistic detector effects for the HE-LHC and FCC-hh.

%
\subsection{ The simulation and detector effects }\label{simulation-detector}
%

In this section, we present the analysis of $p p \to t \bar t t \, (\bar t t \bar t)$ signal process including the FCNC $tqg$, $tq\gamma$, $tqH$ and $tqZ (\sigma^{\mu \nu}, \gamma_{\mu})$ vertices as well as all the relevant SM backgrounds with experimental conditions of the HE-LHC and FCC-hh. For the simulations of the HE-LHC and FCC-hh collider phenomenology, we use the \feynrules~\cite{Alloul:2013bka} to extract the Feynman rules from the effective Lagrangian of Eq.~\eqref{Effective-Lagrangian}. The Universal \feynrules Output ({\tt UFO}) files have been generated~\cite{Degrande:2011ua} and then UFO files fed to the Monte Carlo event generator \madgraph~\cite{Alwall:2014hca,Frederix:2018nkq} to generate the event samples for signal  processes. \madgraph~\cite{Alwall:2014hca,Frederix:2018nkq} is also used to generate background processes. The sample are generated using the leading order (LO) {\tt NNPDF23L01} parton distribution functions (PDFs)~\cite{Ball:2017nwa,Ball:2014uwa,Ball:2012cx} considering the renormalization and factorization scales are set be the threshold value of the
top quark mass, $\mu = \mu_F = \mu_R = m_{\rm top}$. For the parton showering, fragmentation and hadronization of generated signal and backgrounds events we utilized the \pythia~\cite{Sjostrand:2014zea}. During the production of signal and backgrounds samples, all produced jets inside the events forced to be clustered using the \fastjet~\cite{Cacciari:2011ma} considering the anti-$k_t$ jet clustering algorithm with a cone radius of $R = 0.4$~\cite{Cacciari:2008gp}. Finally, we pass all generated events through the  \delphes~\cite{deFavereau:2013fsa}, which handles the detector effect.

We should emphasize here that, for the FCC-hh analysis, we use the default FCC-hh detector card configuration implemented into the \delphes in order to consider the realistic detector effects of the FCC-hh baseline detector. Considering this configuration, the efficiency of $b$-tagging $\epsilon^b (p_T)$, efficiency of $c$-jets $\epsilon^c (p_T)$ and misidentifications rates for the light-jets are assumed to be jet transverse momentum dependent. They are given by~\cite{Oyulmaz:2018irs,Oyulmaz:2019jqr}:

\begin{eqnarray}\label{delphes-card-FCChh}
&&\epsilon^b_{\rm FCC-hh} (p_T) = 0.85 \times (1.0 - p_T ({\text {GeV}})/15000) \nonumber \\ 
&&\epsilon^c_{\rm FCC-hh} (p_T) = 0.05 \times (1.0 - p_T ({\text {GeV}})/15000) \nonumber \\ 
&&\epsilon^{\rm light-jets}_{\rm FCC-hh} (p_T) = 0.01 \times (1.0 - p_T ({\text {GeV}})/15000) \,.
\end{eqnarray}

For the case of HE-LHC projections and in order to produce the Monte Carlo events, we also employed the DELPHES framework for performing a comprehensive high luminosity (HL) CMS detector response simulation. To this end, we have used the HL-LHC detector card configuration implemented into the \delphes  which includes high configuration of the CMS detector~\cite{Azzi:2019yne,CMS:1900mtx,Atlas:2019qfx,Cerri:2018ypt}. The $b$-tagging efficiency $\epsilon^b (p_T)$ and misidentification rates for light-flavor quarks are assumed to be 

\begin{eqnarray}\label{delphes-card-HELHC}
&&\epsilon^b_{\rm HE-LHC} (p_T) = 0.75 \times (1.0 - p_T ({\text {GeV}})/5000) \nonumber \\ 
&&\epsilon^c_{\rm HE-LHC} (p_T) = 0.10 \times (1.0 - p_T ({\text {GeV}})/5000) \nonumber \\ 
&&\epsilon^{\rm light-jets}_{\rm HE-LHC} (p_T) = 0.01 \times (1.0 - p_T ({\text {GeV}})/5000) \,.
\end{eqnarray}

In all numerical calculation, the SM inputs are considered as $m_{\rm top} = 173.34$ GeV for the top quark, $\alpha=1/127.90$ for the coupling constant, $m_{W}$= 80.419 GeV for the $W$ boson mass, $m_{Z}$= 91.187 GeV for the $Z$ boson mass, $\alpha_s(M_Z^2) = 0.1184$ for the strong coupling constant and $s_{W}^{2} = 0.234$~\cite{Tanabashi:2018oca}.

%
\section{  Statistical method for the $tqX$ FCNC analysis }\label{sec:statistical}
%

We detail below the statistical method we assume, together with the numerical calculations and distributions for the HE-LHC and FCC-hh. More details are
provided in the next section. As we discussed in details in Sec.~\ref{sec:signal-and-background}, the studied topology gives rise to the MET + jets + leptons signature characterized by five or more than five jets, and a missing transverse momentum from the undetected neutrino and exactly 2 same-sign isolated charged leptons. Among these jets, three of the them should be tagged as $b$-jets. Based on this signal topology, we follow a standard methodology to distinguish the signal signature from the corresponding SM backgrounds and considered some different preselection cuts as we describe here.

As we applied the leptonic channels of $W$ boson for the top (antitop) quark pairs in the signal process, exactly two same-sign isolated charged leptons  (electron or muon) are required, $n^\ell=2\ell^{\pm\pm}$, with $|\eta^{\ell}|<2.5$ and $p_T^{\ell} > 10$ GeV. As we highlighted before, one of the key ingredients in the strategy pursued in the present study is the triple-top signal with the topology of two same-sign isolated charged leptons. We will discuss in the next section that an additional cut of the same-sign dileptons invariant mass distributions ($M_{\ell^\pm \ell^\pm} > 10$ GeV) need to be taken into account to suppress events with pairs of same-sign energetic leptons from the heavy hadrons decays of backgrounds. 
Since we consider the doubly leptonic decay of $W$ boson in the final state, triple-top signals include a substantial amount of missing transverse energy. For the case of missing transverse energy, we apply $E_T^{\rm miss} > 30$ as a baseline selection~\cite{Oyulmaz:2018irs}. Our signal scenario also includes at least five jets $n^j \geq 5 \, jets$ with $|\eta^{\rm jets}| < 2.5$ and $p_T^{\rm jets} > 20$ GeV. 

We also considered the distance between leading leptons and jets $\Delta R (\ell, j_i) = \sqrt{(\Delta \phi_{\ell, j_i})^2 + (\Delta \eta_{\ell, j_i})^2} > 0.4$ in which are azimuthal angle and the pseudorapidity difference between these two objects. The same selection also need to be taken into account between two jets, $\Delta R (j_i, j_j) > 0.4$. Among the selected jets, at least three of them need to be tagged as $b$ jets, i.e. $n^{b-jets} \geq 3$. 

After adopting our basic cuts and selection of signal and background events, in next section we study the signal and the backgrounds at the level
of distributions and the numerical calculations for the HE-LHC and FCC-hh separately. We should notice here that, in our study that will be discussed in the next section, we only concentrate on the $t \to qg$ and $t \to qH$ modes as a reference throughout this work for presenting some selected distributions. We also choose those distributions which show a good potential to separate the signal for the SM backgrounds.

%
\subsection{ The signal and background analysis and distributions at HE-LHC }\label{Calculations-HE-LHC}
%

After introducing the simulation,  detector effects and the event selection in previous sections, in the this section, we present the numerical calculations and distributions for the HE-LHC scenario.

Let us now present and discuss the cross sections of the triple-top signal and all relevant SM backgrounds in order to provide a basic idea of their production rate. As we discussed before,  the SM four top production ($t \bar t  t   \bar t$),  the SM three top production ($t \bar t t$, $\bar t t \bar t$),  $t \bar t Z$, $t \bar t W$ and $WWZ$ SM backgrounds are the main source of backgrounds considered in this study.

As we mentioned before, the SM three top production ($t \bar t t$, $\bar t t \bar t$) is another important source of backgrounds for the signal signature that we considered in this study. At leading order (LO) in SM, triple-top signal are produced in association with a $b$-jet  ($t \bar t t + \bar b$, $\bar t t \bar t +b$), or a light-jet ($t \bar t t + \text{light-jet}$, $\bar t t \bar t + \text{light-jet}$), or associated with a W boson ($t \bar t t + W^-$, $\bar t t \bar t + W^+$)~\cite{Barger:2010uw,Cao:2019qrb,Chen:2014ewl}.  In this study, we extensively investigated SM three top background for the case of HL-LHC and FCC-hh. The total cross section that we obtained for the $t \bar t t + W^-$,    $\bar t t \bar t + W^+$,  $t \bar t t + j$ and $\bar t t \bar t + j$ with $j=b$, $\bar b$ or a light jet, considering both the leptonic decay ($\ell \nu_{\ell}$) and hadronic decay ($j j$) of $W$ boson,  semileptonic decays of two same-sign top quarks and fully hadronic decay of another top quark, is about $\sigma_{t \bar t t, \bar t t \bar t} = 0.188$ (fb) for the HL-LHC at 27 TeV. 

The SM four top production ($t \bar t  t   \bar t$)~\cite{CMS:1900mtx} considering both semi-leptonic decay and fully hadronic decay of two same-sign top quarks also is examined in this study. The total cross section that we obtained for the four top production ($t \bar t  t   \bar t$), is about $\sigma_{t \bar t  t   \bar t} = 0.940$ (fb) for the HL-LHC at 27 TeV.  We also examine other source of SM backgrounds such as $t \bar t W W$, $t \bar t H$, $t \bar t W W$ and  the top pair $t \bar t$ events in dilepton channel. However, we found that these backgrounds have small contributions in the total background composition.

The cross-sections in the unit of fb for the main SM backgrounds considered in this study including the $t \bar t Z$, $t \bar t W$, $WWZ$,  SM three top production ($t \bar t t$, $\bar t t \bar t$) and SM four top production ($t \bar t  t   \bar t$) passing sequential selection cuts are presented in Table~\ref{tab:Cuts-bkg-27TeV} for the HE-LHC at $\sqrt{s} = 27 \, {\rm TeV}$. As one can see from Table~\ref{tab:Cuts-bkg-27TeV}, these selection criterion could significantly suppress the large contributions of background events originating from the $t\bar{t}Z$ and $t\bar{t}W$, and specially from the $WWZ$, SM three top production ($t \bar t t$, $\bar t t \bar t$) and SM four top production ($t \bar t  t   \bar t$). Among the selection strategy, two same-sign isolated leptons selection $n^{\ell} = 2 \ell^{\pm \pm}$ reduces these backgrounds and lead to the selection efficiencies of 12\%, 37\%, 12\%, 17\% and 28\% for the $t \bar t Z$, $t \bar t W$, $WWZ$, SM three top production and SM four top production, respectively. Selecting jets and $b$-jets, considerably affect all backgrounds as well. For example, the cut efficiency of $b$-jets selection is about 0.38\% for $t \bar t Z$, 0.67\% for $t\bar{t}W$, 0.001\% for $WWZ$, 4.87\% for the SM three top production and 12.8\% for the  SM four top production which all have the same final state with the signal. These small efficiencies indicate that the three tagged $b$-jets selection can reduce the SM backgrounds strongly. The sum of the cross section for all SM backgrounds after all cuts is found to be 0.379 fb. As one can see from Table.~\ref{tab:Cuts-bkg-27TeV}, the contributions from $WWZ$ and SM ($t \bar t t$, $\bar t t \bar t$) are rather small in respect to other backgrounds considered in this study.


\begin{table*}[tbh]
\begin{center}
\begin{tabular}{ c | c c c c c }
$\sqrt {s}$ = 27 TeV (HE-LHC)      ~&~  $t \bar t Z$  ~&~ $t \bar t W$   ~&~ $WWZ$  ~&~ SM ($t \bar t t$, $\bar t t \bar t$) ~&~  SM ($t \bar t  t   \bar t$) \\   \hline  
Cross section (in fb)           &      33.92      &   17.45    &     3.03   &  0.188  &  0.940 \\ 
(I): $n^{\ell}=2\ell^{\pm \pm}$,   $|\eta^{\ell}|<2.5$,   $p_T^{\ell} > 10 \, {\rm GeV}$, $M_{\ell^\pm \ell^\pm} > 10 \, {\rm GeV}$     & 4.26   & 6.57 &  0.36    &  0.032  & 0.271 \\
(II): $\missE >  30 \, {\rm GeV}$       & 3.62    & 5.90  &  0.28    &  0.029    & 0.251  \\
(III): $n^{\rm jets} \geq 5 \, {\rm jets}$,   $|\eta^{\rm jets}|<2.5$,   $p_T^{\rm jets} > 20 \, {\rm GeV}$,   $\Delta R(\ell, j_i) \geq 0.4$,   $\Delta R(j_i, j_j) \geq 0.4$      &  1.31  &  1.06  &   0.011     &   0.022 &  0.227   \\	
(IV): $n^{\rm b-jets}  \geq 3 \, b \, {\rm jets}$    & 0.132 	& 0.117 &  0.00003    &   0.009  &  0.121 \\ 	\hline  \hline
\end{tabular}
\end{center}
\caption{ Cross-sections in the unit of fb for the  main SM backgrounds considered in this study including the $t \bar t Z$, $t \bar t W$, $WWZ$ and the SM three top production ($t \bar t t$, $\bar t t \bar t$) passing sequential selection cuts at HE-LHC collider.  }
\label{tab:Cuts-bkg-27TeV}
\end{table*}

After our discussion on the numerical calculations for the backgrounds, let us now present our signal calculations. We should notice here that, the
fixed values of $\zeta_{qt}=1$, $\eta_{qt}=1$, $\kappa_{qt}=1$, $X_{qt}=1$ and $\lambda_{qt}=1$ with $q=u, c$ are chosen for all coupling strength as the benchmark point if not stated otherwise. Taking these typical benchmarks input for the triple-top signal, the expected cross sections before and after the selection cuts are presented in Table~\ref{tab:Cuts-sig-27TeV}.

\begin{table*}[tbh]
\begin{center}
\begin{tabular}{c|c|c|c|c|c}
$\sqrt {s}$ = 27 TeV  (HE-LHC)        ~&~  Cross section (in fb)           ~&~  Cut (I) ~&~  Cut (II) ~&~  Cut (III) ~&~ Cut (IV)   \\      \hline     \hline
$tug $                     & $2.48 \times 10^7 \, (\zeta_{tu})^{2}$    & $5.71 \times 10^6  \, (\zeta_{tu})^{2}$ & $5.51 \times 10^6  \, (\zeta_{tu})^{2}$ & $3.78 \times 10^6  \, (\zeta_{tu})^{2}$ & $1.53 \times 10^6 \, (\zeta_{tu})^{2}$  \\
$tcg $                     & $2.19 \times 10^6  \, (\zeta_{tc})^{2}$    & $6.58 \times 10^5   \, (\zeta_{tc})^{2}$ & $6.30 \times 10^5  \, (\zeta_{tc})^{2}$ & $4.21 \times 10^5   \, (\zeta_{tc})^{2}$ & $1.74 \times 10^5 \, (\zeta_{tc})^{2}$  \\
$tuH$                       & $43.53  \, (\eta_{tu})^{2}$     & $15.57   \, (\eta_{tu})^{2}$  & $14.18   \, (\eta_{tu})^{2}$  & $7.06   \, (\eta_{tu})^{2}$ & $2.72   \, (\eta_{tu})^{2}$  \\ 
$tcH$                       & $8.18   \, (\eta_{tc})^{2}$     & $3.18   \, (\eta_{tc})^{2}$  & $2.88   \, (\eta_{tc})^{2}$ & $1.37   \, (\eta_{tc})^{2}$ & $0.52   \, (\eta_{tc})^{2}$  \\	
$tu\gamma $                 & $43.74   \, (\lambda_{tu})^{2}$  &   $14.09   \, (\lambda_{tu})^{2}$  & $13.18   \, (\lambda_{tu})^{2}$ & $7.42   \, (\lambda_{tu})^{2}$ & $2.94   \, (\lambda_{tu})^{2}$    \\
$tc\gamma $                 & $6.19  \, (\lambda_{tc})^{2}$  &   $2.28   \, (\lambda_{tc})^{2}$  & $2.12   \, (\lambda_{tc})^{2}$ &  $1.15   \, (\lambda_{tc})^{2}$ & $0.45   \, (\lambda_{tc})^{2}$ \\
$tuZ$    $(\sigma_{\mu\nu})$ & $295.99 \, (\kappa_{tu})^{2}$   &   $91.83 \, (\kappa_{tu})^{2}$   & $85.41 \, (\kappa_{tu})^{2}$  & $47.12 \, (\kappa_{tu})^{2}$ & $18.88 \, (\kappa_{tu})^{2}$  \\

$tcZ$    $(\sigma_{\mu\nu})$ & $42.17 \, (\kappa_{tc})^{2}$   &   $15.13 \, (\kappa_{tc})^{2}$ &  $13.98 \, (\kappa_{tc})^{2}$  & $7.49 \, (\kappa_{tc})^{2}$ & $2.94 \, (\kappa_{tc})^{2}$ \\

$tuZ$    $(\gamma_{\mu})$  & $152.56   \, (X_{tu})^{2}$        &  $47.77   \, (X_{tu})^{2}$    &  $43.02   \, (X_{tu})^{2}$  & $20.04   \, (X_{tu})^{2}$   & $7.57   \, (X_{tu})^{2}$  \\ 

$tcZ$    $(\gamma_{\mu})$  & $27.57   \, (X_{tc})^{2}$        &  $10.11   \, (X_{tc})^{2}$    &  $9.08   \, (X_{tc})^{2}$ & $4.13   \, (X_{tc})^{2}$   & $1.55   \, (X_{tc})^{2}$ \\ 	\hline  \hline
\end{tabular}
\end{center}
\caption{   Cross-sections in the unit of fb for the triple-top production at HE-LHC  $p p \to t  \bar t t \, (\bar t t \bar t)$ with $\ell = e, \mu$ for five signal topologies of $tqg$, $tqH$, $tqZ (\sigma_{\mu\nu})$, $tqZ (\gamma_{\mu})$ and $tq\gamma$ before and after passing sequential selection cuts.   }
\label{tab:Cuts-sig-27TeV}
\end{table*}

The following conclusions can be drawn from the cut flow table present in Table~\ref{tab:Cuts-sig-27TeV}. For all triple-top signal topologies, around 35\% efficiency obtained considering the same-sign dilepton selection and 12-15\% efficiency achieved after the jet selection strategy. As one can see, after the sample selections, around 6-7\% of signal events could pass the selection criteria.

\begin{figure*}[htb]
\begin{center}
\vspace{0.50cm}
\resizebox{0.480\textwidth}{!}{\includegraphics{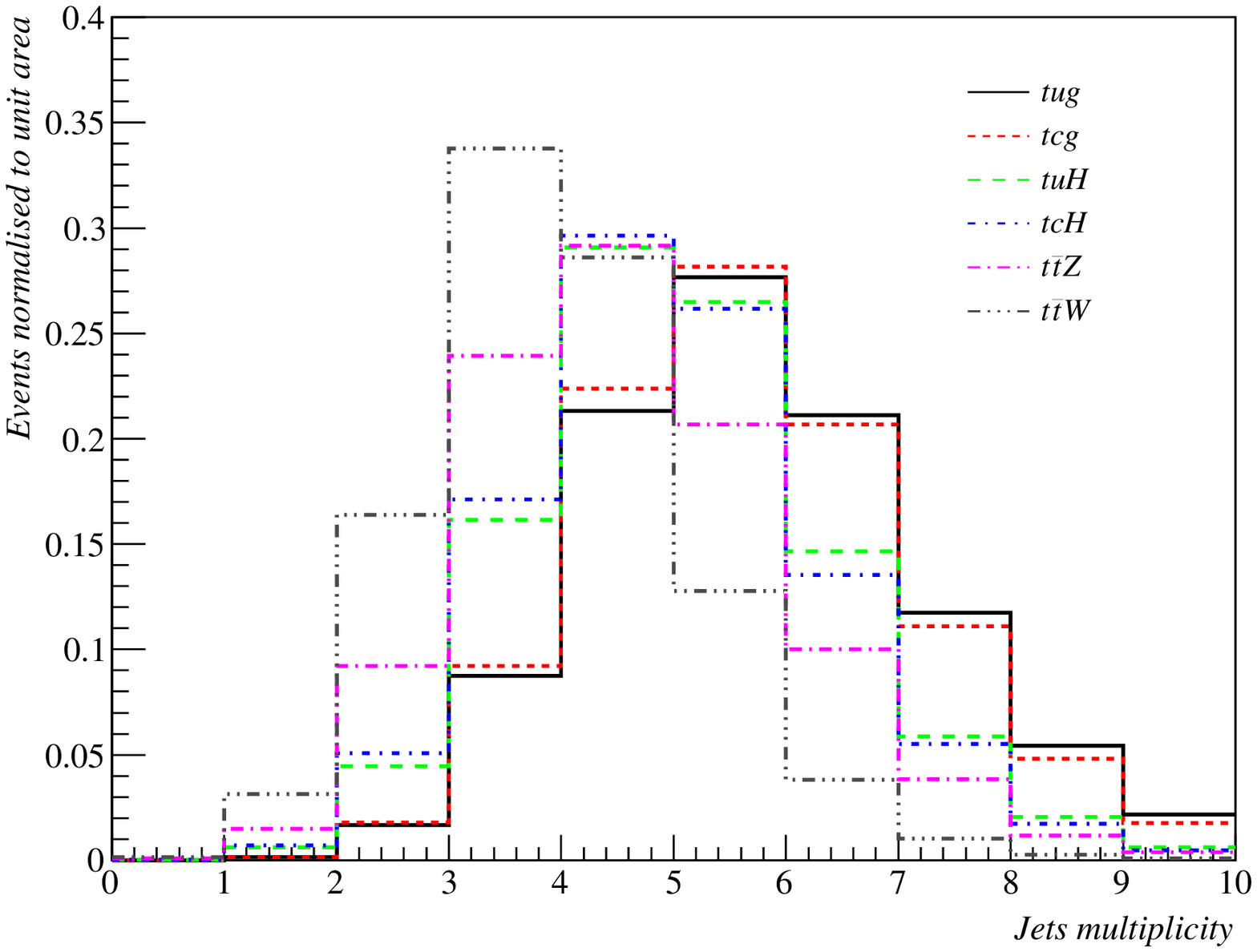}}   	
\resizebox{0.480\textwidth}{!}{\includegraphics{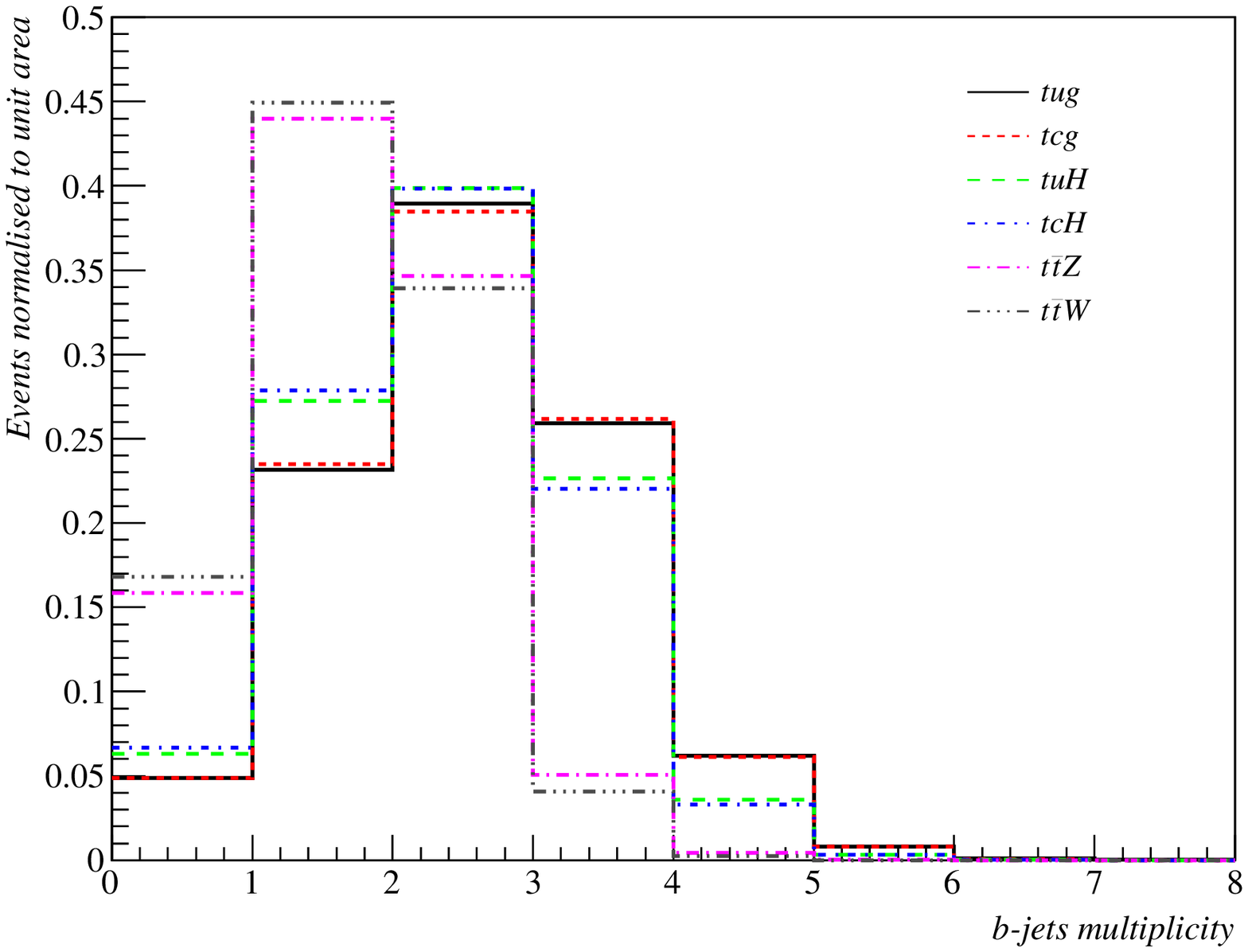}}  	
\caption{ Jets multiplicity distributions (left) and $b$-jets multiplicity distributions (right) for the triple-top signal of $tqg$ and $tqH$ obtained form \madgraph~\cite{Alwall:2014hca,Frederix:2018nkq} at leading order for HE-LHC at 27 TeV. The main SM backgrounds $t \bar t Z$ and $t \bar t W$ also are presented as well. } \label{fig:Multi-HE-LHC}
\end{center}
\end{figure*}

Now let us discuss the triple-top signal and the SM backgrounds at the distribution level. Characteristic signature of the triple-top signal process analyzed in this study suggests to work with the events having at least two isolated same-sign
lepton in which can be an electron or a muon, large missing transverse energy (MET), and at least 5 jets which three of them are required to be identified as jets originating from the $b$-quark. Considering these signal scenarios and all relevant SM backgrounds, in Fig.~\ref{fig:Multi-HE-LHC}, we show the jets and $b$-jets multiplicities for triple-top signal and the main SM backgrounds events before applying the jet and $b$-jet selection. All the plots are unit normalized. As we mentioned before, we only concentrate on the $t \to q g$ and $t \to q H$ modes as a reference throughout this work for presenting some selected distributions. Hence, for the signal in these figures, only one coupling ($\zeta_{tq}$ or $\eta_{tq}$ with $q=u,c$) at a time is varied from its SM value. It can be concluded from these plots, the requirement of at least five jets ($n^{\rm jets} \geq 5$) is useful to reduce the contributions of the SM backgrounds. In addition to this requirement, selecting at least three $b$-tagged jets $n^{\rm b-jets} \geq 3$ among those jets is also useful to suppress the contribution of the SM backgrounds.

%
\subsection{ The signal and background analysis and distributions at FCC-hh } \label{Calculations-FCC-hh}
%

In this section, we focus on the numerical calculations of the FCC-hh collider for the triple-top signal analyzed in this study. 
As we mentioned before, in addition to the HE-LHC, we also plan to investigate the potential of future FCC-hh collider to the top quark FCNC couplings at a center of mass energy of 100 TeV and present our study to set an upper limits on the anomalous top FCNC  $tqX (X = g, H, Z, \gamma)$ vertices including the realistic detector effects. However, one can even expect the higher center-of-mass energies of FCC-hh collider could lead to the improvement of these limits. For the FCC-hh analysis, we follow the same strategy applied for the case of HE-LHC, namely the the semileptonic final state with two same-charged $W$ decaying to either electron or muon, and the other decaying hadronically. We also use the same selection cuts for the signal and backgrounds.  We expect, and do find, a similar peak on the jets and $b$-jets multiplicity distributions for the signal and for the backgrounds as the case of HE-LHC presented in Fig.~\ref{fig:Multi-HE-LHC}. The difference between the HE-LHC and FCC-hh mainly comes from the different detector configurations. As we discussed in Section~\ref{simulation-detector}, for the FCC-hh analysis, we use the default FCC-hh detector card configuration implemented into the \delphes in order to consider the realistic detector effects of the FCC-hh baseline detector. All the SM backgrounds listed above for the HE-LHC are also belonging to the backgrounds of FCC-hh.

Notice that while the colliding energy of FCC-hh is different, we consider the same selection cuts which are optimized for the HE-LHC. We find that these selections also reasonable for the case of FCC-hh.  As we mentioned before, we consider the SM three top production ($t \bar t t$, $\bar t t \bar t$) and SM four top production ($t \bar t  t   \bar t$) as important source of backgrounds. For the case of FCC-hh energy, the cross section for the SM three top production ($t \bar t t + \bar b$,       $\bar t t \bar t +b$,      $t \bar t t + \text{light-jets}$,     $\bar t t \bar t + \text{light-jets}$,   $t \bar t t + W^-$,   $\bar t t \bar t + W^+$) considering both the leptonic decay ($\ell \nu_{\ell}$) and hadronic decay ($j j$) of $W$ boson, semileptonic decays of two same-sign top quarks and fully hadronic decay of another top quark is estimated to be around $\sigma_{t \bar t t, \bar t t \bar t} = 2.83$ fb which is much larger than the case of HE-LHC. The cross section for the SM four top production ($t \bar t  t   \bar t$) with semi-leptonic decay and fully hadronic decay of two same-sign top quarks is estimated to be around $\sigma_{t \bar t  t   \bar t} = 15.18$ fb which is also much larger than the case of HE-LHC.

Hence, the contribution of these two SM backgrounds need to be taken into account in our FCC-hh study. Considering these points, in Table~\ref{tab:Cuts-bkg-100TeV}, we present the cross section measurements of the SM backgrounds at FCC-hh before and after applying the selection cuts.

As one can see, selection of two same-sign isolated charged lepton could significantly affects the $t \bar t Z$ and $WWZ$ backgrounds with 11\% of efficiency and  the SM three top production with 19\% of efficiency. This selection also leads to 43\% for the $t \bar t W$ backgrounds. This finding indicates that, among all the backgrounds, the $t \bar t W$ is indeed hard to suppress. One can see that, the jets selection also affects the mentioned backgrounds considerably. Finally, our selection strategy leads to 0.8\% of $t \bar t Z$, 2\% of $t \bar t W$ and 0.004\% of $WWZ$ backgrounds, 8.9\% of the SM three top production and 21.5\% for the SM four top production ($t \bar t  t   \bar t$). The sum of the cross section for all SM backgrounds after all cuts is found to be 7.81 fb.

\begin{table*}[tbh]
\begin{center}
\begin{tabular}{ c | c c c c c }
$\sqrt {s}$ = 100 TeV  (FCC-hh)       ~&~  $t \bar t Z$  ~&~ $t \bar t W$   ~&~ $WWZ$  	~&~ SM ($t \bar t t$, $\bar t t \bar t$)  ~&~ SM ($t \bar t  t   \bar t$) \\   \hline  
Cross section (in fb)           &      343.44      &   73.16    &     12.45   &  2.83  &   15.18    \\ 
(I): $n^{\ell}=2\ell^{\pm \pm}$,   $|\eta^{\ell}|<2.5$,   $p_T^{\ell} > 10 \, {\rm GeV}$, $M_{\ell^\pm \ell^\pm} > 10 \, {\rm GeV}$    & 39.424   & 32.09 &  1.30  &  0.54 &  5.03   \\
(II): $\missE >  30 \, {\rm GeV}$       & 33.23    & 28.96  &  1.01   &  0.50  &  4.68  \\
(III): $n^{\rm jets} \geq 5 \, {\rm jets}$,   $|\eta^{\rm jets}|<2.5$,   $p_T^{\rm jets} > 20 \, {\rm GeV}$,   $\Delta R(\ell, j_i) \geq 0.4$,   $\Delta R(j_i, j_j) \geq 0.4$      &  19.59  &  11.53  &   0.11    &   0.46   &  4.60   \\	
(IV): $n^{\rm b-jets}  \geq 3 \, b \, {\rm jets}$    & 2.76 	& 1.53 &  0.00056   &  0.25   & 3.27   \\    	\hline    \hline
\end{tabular}
\end{center}
\caption{ Cross-sections in the unit of fb  for the $t \bar t Z$, $t \bar t W$ and $WWZ$ SM backgrounds, and  SM three top production ($t \bar t t$, $\bar t t \bar t$)  passing sequential selection cuts for the FCC-hh collider.  }
\label{tab:Cuts-bkg-100TeV}
\end{table*}

Our numerical calculations of the triple-top signal cross sections at leading order in the presence of the top quark FCNC vertices are presented in details in Table~\ref{tab:Cuts-sig-100TeV}. This table shows the cut flow dependence of various signal scenarios studied here. One clearly sees that we achieved to the selection efficiency of 6\% for $tug$, 10\% for $tcg$ and around 15\% for all other signal scenarios. 

%
\begin{table*}[tbh]
\begin{center}
\begin{tabular}{c|c|c|c|c|c}
$\sqrt {s}$ = 100 TeV (FCC-hh)      ~&~  Cross section (in fb)           ~&~  Cut (I) ~&~  Cut (II) ~&~  Cut (III) ~&~ Cut (IV)   \\      \hline     \hline
$tug $       & $3.64  \times 10^8 \, (\zeta_{tu})^{2}$    & $5.39  \times 10^7  \, (\zeta_{tu})^{2}$ & $5.23  \times 10^7  \, (\zeta_{tu})^{2}$ & $4.38 \times 10^7  \, (\zeta_{tu})^{2}$ & $2.19 \times 10^7 \, (\zeta_{tu})^{2}$  \\
$tcg $        & $5.91  \times 10^7 \, (\zeta_{tc})^{2}$    & $1.48   \times 10^7 \, (\zeta_{tc})^{2}$ & $1.43   \times 10^7 \, (\zeta_{tc})^{2}$ & $1.19   \times 10^7 \, (\zeta_{tc})^{2}$ & $6.15   \times 10^6 \, (\zeta_{tc})^{2}$  \\
$tuH$       & $212.10   \, (\eta_{tu})^{2}$     & $91.25   \, (\eta_{tu})^{2}$  & $83.41   \, (\eta_{tu})^{2}$  & $62.23   \, (\eta_{tu})^{2}$ & $31.94   \, (\eta_{tu})^{2}$  \\ 
$tcH$      & $86.46   \, (\eta_{tc})^{2}$     & $39.17   \, (\eta_{tc})^{2}$  & $35.69   \, (\eta_{tc})^{2}$ & $26.57   \, (\eta_{tc})^{2}$ & $13.59   \, (\eta_{tc})^{2}$  \\
$tu\gamma $                 & $266.24   \, (\lambda_{tu})^{2}$  &   $95.11   \, (\lambda_{tu})^{2}$  & $89.51   \, (\lambda_{tu})^{2}$ & $69.29   \, (\lambda_{tu})^{2}$ & $35.41   \, (\lambda_{tu})^{2}$    \\
$tc\gamma $                 & $85.49   \, (\lambda_{tc})^{2}$  &   $34.95   \, (\lambda_{tc})^{2}$  & $32.68   \, (\lambda_{tc})^{2}$ &  $25.28   \, (\lambda_{tc})^{2}$ & $12.96   \, (\lambda_{tc})^{2}$ \\				
$tuZ$    $(\sigma_{\mu\nu})$ & $1732.59 \, (\kappa_{tu})^{2}$   &   $608.23 \, (\kappa_{tu})^{2}$   & $568.52 \, (\kappa_{tu})^{2}$  & $438.28 \, (\kappa_{tu})^{2}$ & $227.39 \, (\kappa_{tu})^{2}$  \\
$tcZ$    $(\sigma_{\mu\nu})$ & $566.91 \, (\kappa_{tc})^{2}$   &   $229.32 \, (\kappa_{tc})^{2}$ &  $213.56 \, (\kappa_{tc})^{2}$  & $164.52 \, (\kappa_{tc})^{2}$ & $86.13 \, (\kappa_{tc})^{2}$  \\
$tuZ$    $(\gamma_{\mu})$  & $740.51   \, (X_{tu})^{2}$        &  $285.98   \, (X_{tu})^{2}$    &  $258.52   \, (X_{tu})^{2}$  & $189.26   \, (X_{tu})^{2}$   & $97.35   \, (X_{tu})^{2}$  \\ 
$tcZ$    $(\gamma_{\mu})$  & $293.83   \, (X_{tc})^{2}$        &  $125.00   \, (X_{tc})^{2}$    &  $112.99   \, (X_{tc})^{2}$ & $82.65   \, (X_{tc})^{2}$   & $41.95   \, (X_{tc})^{2}$ \\    \hline    \hline
\end{tabular}
\end{center}
\caption{   Cross-sections in the unit of fb for the triple-top production at FCC-hh  $p p \to t \bar t t \, (\bar t t \bar t)$ with $\ell = e, \mu$ for five signal topologies of $tqg$, $tqH$, $tqZ (\sigma_{\mu\nu})$, $tqZ (\gamma_{\mu})$ and $tq\gamma$ before and after passing sequential preselection cuts.  }
\label{tab:Cuts-sig-100TeV}
\end{table*}
%

\begin{figure*}[htb]
\begin{center}
\vspace{0.50cm}
\resizebox{0.480\textwidth}{!}{\includegraphics{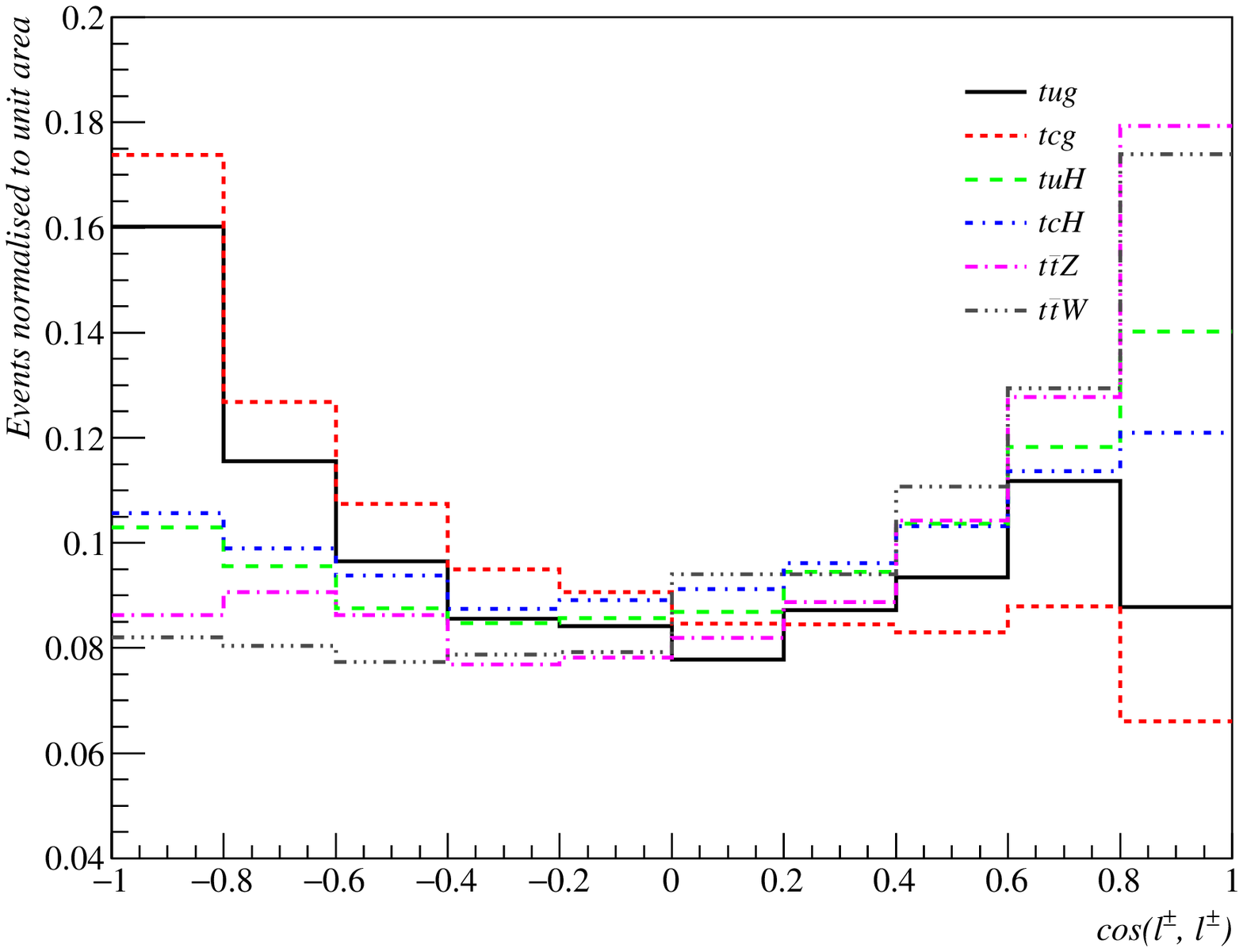}}   		
\resizebox{0.480\textwidth}{!}{\includegraphics{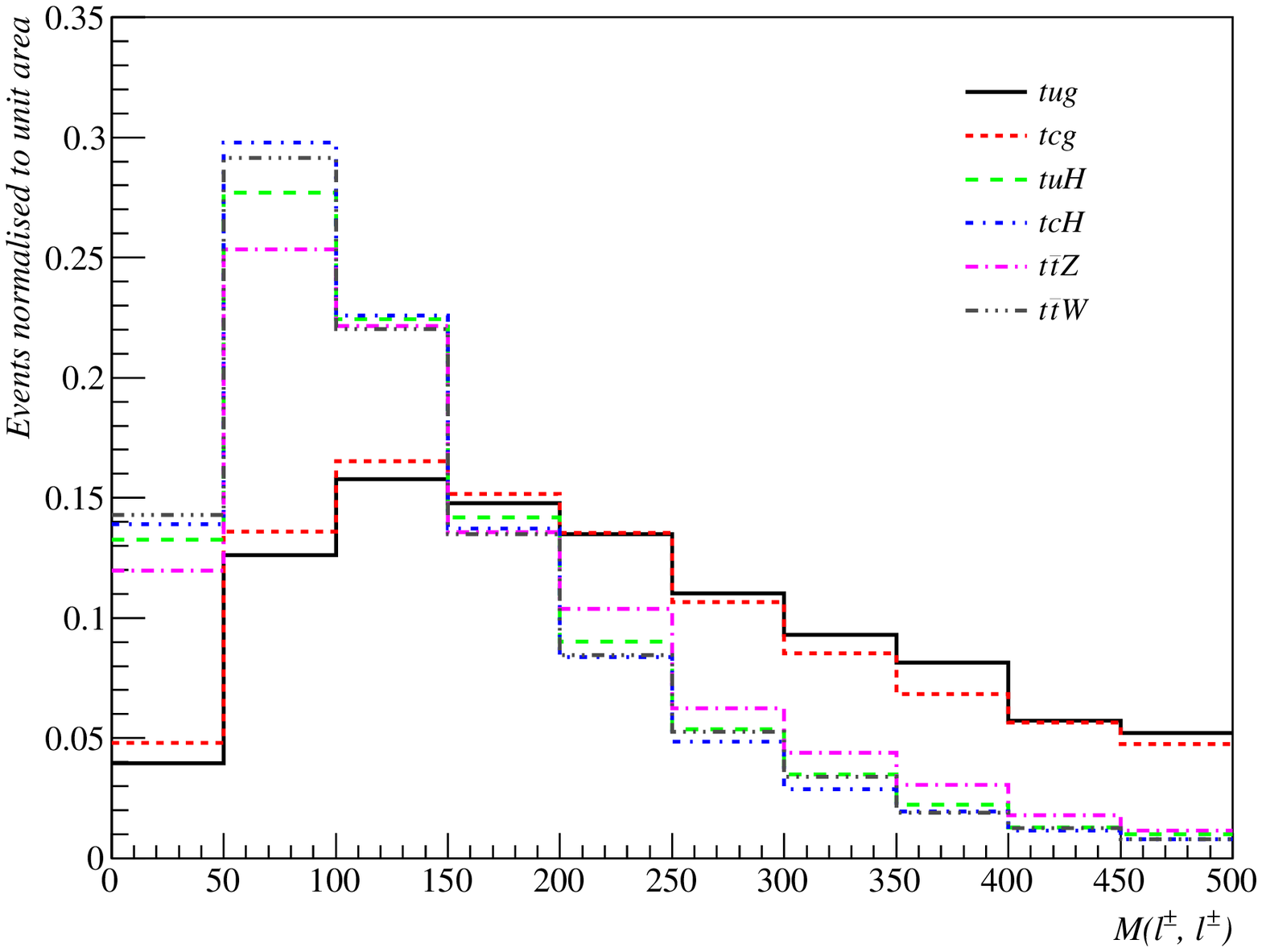}}   		
\caption{ The cosine between two same-sign  leptons $\cos (\ell^\pm, \ell^\pm)$ (left) and the invariant mass distributions of dilepton $M_{\ell^\pm \ell^\pm}$ (right) for $tqg$ and $tqH$ signal scenarios obtained from \madgraph~\cite{Alwall:2014hca,Frederix:2018nkq} at leading order for FCC-hh at 100 TeV. The main SM backgrounds $t \bar t Z$ and $t \bar t W$ also are presented as well. } \label{fig:cosll-Mll-FCChh}
\end{center}
\end{figure*}

For completeness, we also depict in Fig.~\ref{fig:cosll-Mll-FCChh}, some selected distributions, including the cosine between two same-sign  leptons $\cos (\ell^\pm, \ell^\pm)$ (left) and the invariant mass distributions of dilepton $M_{\ell^\pm \ell^\pm}$ (right) for $tqg$ and $tqH$ signal scenarios and the corresponding $t \bar t Z$ and $t \bar t W$ SM backgrounds for the FCC-hh collider. As we explained before, an additional cut on the same-sign dileptons invariant mass distributions ($M_{\ell^\pm \ell^\pm} > 10$ GeV) has been applied to suppress events with pairs of same-sign energetic leptons from the heavy hadrons decays of backgrounds.

The different shapes of $tqg$ couplings in Fig.~\ref{fig:cosll-Mll-FCChh} from other FCNC couplings and also the main SM backgrounds deserve further detailed explanations.
As we already mentioned and discussed in Sec.~\ref{sec:framework}, the triple-top signal containing $tqg$ anomalous FCNC vertex is directly related to the gluon and $u$ ($c$)-quark PDFs.   
In addition, the Feynman diagrams for the FCNC triple-top quark production presented in Fig.~\ref{fig:Feynman-Sginal-1} show both $tqg$ vertex in production and decay, 
which is not the case for the other FCNC couplings.
As a consequence, as can be seen in the right plot of Fig.~\ref{fig:cosll-Mll-FCChh}, one would expect that the final state would be more energetic in the $tqg$ case with respect to the others.  A comparison of the cosine of the angle between the same-sign leptons $\cos (\ell^\pm, \ell^\pm)$ and invariant mass $M_{\ell^\pm \ell^\pm}$ of them point to the fact that one could distinguish between the $tqg$ FCNC couplings and the other ones.

%
\section{  Analysis results and 95\% confidence level limits at HE-LHC and FCC-hh colliders }\label{sec:results}
%

In this section, we present the main results of this study  with detailed discussions and proper explanations,  namely the upper limits on the coupling strengths obtained from the fast simulation of the triple-top signal at HE-LHC and FCC-hh. First, we discuss the upper limits obtained in this study focusing on the  95\% CL limits of the HE-LHC and FCC-hh, and then we compare the results with other studies in literature. Finally, we detail a number of updates and improvements that should be foreseen for the future work. 
Considering the optimized selections of signal and backgrounds that we discussed in section~\ref{sec:statistical} and having at hand the signal efficiencies and the number of backgrounds, we set 95\% CL upper limits on the anomalous FCNC couplings and determine the expected limits on the FCNC branching fractions $Br (t \to qX)$ using a Bayesian approach~\cite{cla}.

There are several sources of uncertainties that one needs to be taken into account to determine the 95\% CL upper limits. The statistical uncertainty in the signal efficiency and the background estimate and the uncertainty in the integrated luminosity is considered in this study. A more realistic consideration of uncertainties could affect the obtained sensitivities. There are other sources of theoretical and statistical uncertainties such as the uncertainty associated with the choice of Parton Distribution Functions (PDFs), uncertainty coming from the variation of renormalization and factorization scales ($\mu_R$ and $\mu_F$), the uncertainty of strong coupling constant $\alpha_s(M_Z)$, and the systematic uncertainty on the main backgrounds, etc. could be considered to set the 95\% CL upper limits~\cite{CMS:2019too,Sirunyan:2017uzs}. The impact of these sources of uncertainties could be estimated, and certainty will weaken the limits. More recently, CMS Collaboration at CERN measured the top quark pair production in association with a $Z$ boson in $pp$ collisions at the center-of-mass energy of 13 TeV. The production cross-section is measured to be $\sigma (t \bar t Z)$ = 0.95 $\pm$ 0.05 (stat) $\pm$ 0.06 (syst) pb~\cite{CMS:2019too}. In addition, the cross-section for top quark pair production in association with a $W$ or $Z$ boson also measured by the same Collaboration at the same center-of-mass energy. The $t\bar tW$ and  $t \bar t Z$  production cross sections are measured to be $\sigma(t\bar tW)= 0.77^{+0.12}_{-0.11} ({\text stat})^{+0.13}_{-0.12} ({\text syst})  \ ({\text pb})$ and $\sigma(t \bar t Z)=0.99^{+0.09}_{-0.08} ({\text stat})^{+0.12}_{-0.10} ({\text syst}) \ ({\text pb})$. The statistical and systematic uncertainties reported in these experiments could be considered as a source of uncertainties for the determination of 95\% CL upper limits in this study. However, one expected that the future colliders could measure the cross-section of these backgrounds with much smaller uncertainties.

The 95\% CL constraints on various FCNC branching fractions including the gluon, Higgs boson, Z boson and photon are detailed and summarized in Table~\ref{limits-u-c-27-10-15-20ab-1} for HE-LHC working at three scenarios of integrated luminosities of ${\cal L}_{int} = 10,\, 15$ and 20 ab$^{-1}$ of data. The most recent experimental constraints on the corresponding branching ratio of the top quark FCNC transitions obtained at the ATLAS and CMS with 95\% CL are also presented as well~\cite{Khachatryan:2016sib,Aaboud:2018pob,Khachatryan:2015att,CMS:2017twu}.

\begin{table*}[htbp]
\begin{center}
\begin{tabular}{c c c c c c }  \hline
Branching fraction    ~&~    HE-LHC 10 ab$^{-1}$     ~&~   HE-LHC 15 ab$^{-1}$       ~&~   HE-LHC 20 ab$^{-1}$  ~&~   LHC (CMS \& ATLAS)      \\    \hline  \hline
$Br (t \to u g)$  &   $8.39 \times 10^{-8}$   &    $6.52 \times 10^{-8}$   &  $5.49 \times 10^{-8}$  &  $2.0 \times 10^{-5}$~\cite{Khachatryan:2016sib} 	\\ 
$Br (t \to u H)$  &   $9.98 \times 10^{-5}$  &    $7.58 \times 10^{-5}$   &  $5.99 \times 10^{-5}$   &  $1.90 \times 10^{-3}$~\cite{Aaboud:2018pob}  \\
$Br (t \to u \gamma)$  &     $2.22 \times 10^{-3}$  &    $1.72 \times 10^{-3}$   &    $1.45 \times 10^{-3}$  &    $1.30 \times 10^{-4}$~\cite{Khachatryan:2015att} \\
$Br (t \to u Z)$  $(\sigma_{\mu \nu})$ &   $3.09 \times 10^{-4}$  &   $2.40 \times 10^{-4}$   &    $2.02 \times 10^{-4}$   &    $1.50 \times 10^{-4}$~\cite{CMS:2017twu} \\
$Br (t \to u Z)$  $(\gamma_{\mu})$ &   $1.07 \times 10^{-3}$   &   $8.36 \times 10^{-4}$   &    $7.02 \times 10^{-4}$  &    $3.64 \times 10^{-4}$ \\  \hline  \hline
$Br (t \to c g)$  &   $7.52 \times 10^{-7}$   &    $5.85 \times 10^{-7}$   &   $4.93 \times 10^{-7}$ 	 &   $4.1 \times 10^{-4}$~\cite{Khachatryan:2016sib}  \\ 
$Br (t \to c H)$  &   $9.00 \times 10^{-4}$  &    $7.01 \times 10^{-4}$   &   $5.88 \times 10^{-4}$   &   $1.60 \times 10^{-3}$~\cite{Aaboud:2018pob}  \\
$Br (t \to c \gamma)$  &     $1.49 \times 10^{-2}$  &    $1.15 \times 10^{-2}$   &   $9.77 \times 10^{-3}$  &   $1.70 \times 10^{-3}$~\cite{Khachatryan:2015att} \\
$Br (t \to c Z)$  $(\sigma_{\mu \nu})$ &   $2.01 \times 10^{-3}$  &   $1.56 \times 10^{-3}$   &  $1.31 \times 10^{-3}$    &  $3.70 \times 10^{-4}$\cite{CMS:2017twu}    \\
$Br (t \to c Z)$  $(\gamma_{\mu})$ &   $5.38 \times 10^{-3}$   &   $4.19 \times 10^{-3}$   &   $3.35 \times 10^{-3}$   &   $1.76 \times 10^{-3}$ \\   \hline  \hline
\end{tabular}
\end{center}
\caption{  The upper limits on the $tuX$ and $tcX$  FCNC at 95\% CL obtained at the 27 TeV HE-LHC for three scenarios of integrated luminosities  ${\cal L} = 10,\, 15$ and 20 ${\rm ab}^{-1}$ of data. The most recent experimental 95\% CL upper limits on the branching fractions of the top quark FCNC transitions obtained at the LHC experiments also have been shown for comparisons.   }
\label{limits-u-c-27-10-15-20ab-1}
\end{table*}

A few remarks concerning the results presented in this table are in order. As one can see, with an integrated luminosity of 15 ab$^{-1}$, the sensitivity to the branching ratio of $tug$ and $tcg$ channels can be constrained to the order of ${\cal O} (10^{-6})$\%  and ${\cal O} (10^{-5})$\% respectively, which are three order of magnitude better than the available experimental limits form CMS Collaboration~\cite{Khachatryan:2016sib}. For the $tuH$ and $tcH$ we obtained ${\cal O} (10^{-3})$\%  and ${\cal O} (10^{-2})$\% respectively, showing that the limits calculated in this study are two and one orders to magnitude better than the most recent direct limits on the corresponding branching ratios reported by ATLAS Collaborations at CERN with an integrated luminosity of 36.1 fb$^{-1}$ at $\sqrt{s} = 13 \, {\rm TeV}$. We should notice here that the limits by ATLAS and CMS presented in Table.~\ref{limits-u-c-27-10-15-20ab-1} have measured for the different final state of proton-proton collision and different center-of-mass energies and luminosities. These limits are discussed in details in the Introductions. As a short summary and based on the HE-LHC projections, one can conclude that in the cases of $tqg$ and $tqH$ FCNC transitions, the most stringent constraints have been obtained, and hence, the limits through the tree-top signal can be much better than other channel studied in the literature. 
For other FCNC transitions, namely $tq\gamma$ and $tqZ$, we determined comparable branching ratios with the recent experimental limits~\cite{Khachatryan:2015att,CMS:2017twu}. This finding indicates that the tree top signal analyzed in this study is not much sensitive to these particular FCNC vertices.

At this stage, we turn to present and discuss upper limits on the signal rates at 95\% CL for the case of FCC-hh scenario. We exactly follow the statistical method used for the HE-LHC study to set upper limits on the coupling strengths and the resulting branching fractions. The branching fractions $Br (t \to qX)$  for the FCC-hh at the luminosity of 10 ab$^{-1}$ are detailed in Table~\ref{limits-u-c-100-10ab-1}. The results of our recent top-quark FCNC study considering triple-top signal at 14 TeV LHC have been presented as well~\cite{Malekhosseini:2018fgp}.

\begin{table}[htbp]
\begin{center}
\begin{tabular}{c c c c c c c }  \hline
Branching fraction    &    FCC-hh 10 ab$^{-1}$   &    LHC 14 TeV, 3 ab$^{-1}$~\cite{Malekhosseini:2018fgp}    \\    \hline  \hline
$Br (t \to u g)$  &   $2.04 \times 10^{-8}$ &  $1.19 \times 10^{-4}$	\\ 
$Br (t \to u H)$  &   $4.79 \times 10^{-5}$  & $3.09 \times 10^{-4}$  \\
$Br (t \to u \gamma )$  &     $6.26 \times 10^{-4}$ & $6.53 \times 10^{-3}$  \\
$Br (t \to u Z)$  $(\sigma_{\mu \nu})$ &   $8.65 \times 10^{-5}$ & $8.18 \times 10^{-4}$  \\
$Br (t \to u Z)$  $(\gamma_{\mu})$ &   $2.76 \times 10^{-4}$   &    $1.71 \times 10^{-3}$    \\   \hline  \hline
$Br (t \to c g)$  &   $7.27 \times 10^{-8}$  & $1.35 \times 10^{-3}$	\\ 
$Br (t \to c H)$  &   $9.68 \times 10^{-5}$  &  $2.54 \times 10^{-3}$  \\
$Br (t \to c \gamma )$  &     $1.74 \times 10^{-3}$ &  $6.40 \times 10^{-2}$  \\ 
$Br (t \to c Z)$  $(\sigma_{\mu \nu})$ &   $2.33 \times 10^{-4}$ &  $7.98 \times 10^{-3}$    \\ 
$Br (t \to c Z)$  $(\gamma_{\mu})$ &   $6.52 \times 10^{-4}$   & $1.35 \times 10^{-2}$   \\   \hline  \hline
\end{tabular}
\end{center}
\caption{  The upper limits on the $tuX$ and $tcX$  FCNC at 95\% CL obtained at the 100 TeV FCC-hh for the integrated luminosity  of ${\cal L} = 10$ ${\rm ab}^{-1}$ of data. The results of our recent top-quark FCNC study for tree top production at 14 TeV LHC have been presented as well~\cite{Malekhosseini:2018fgp}.   }
\label{limits-u-c-100-10ab-1}
\end{table}

Similar conclusions as those for the HE-LHC can be drawn for the FCC-hh. However, one can expect further improvements with a higher center-of-mass energy collider. Our study suggests that improvement on the upper limits of all analyzed FCNC couplings can be obtained at FCC-hh collider; which is found to be around 80\% for the $tug$ and one order of magnitude for the $tcg$  coupling with respect to the HE-LHC. The improvements on these bounds are likely due to the higher center of mass energy of FCC-hh. In comparison to the results of our recent top-quark FCNC study considering triple-top signal at 14 TeV LHC~\cite{Malekhosseini:2018fgp}, one can see that remarkable improvements have been obtained for the $tqg$ couplings. As one can see from Tabl.~\ref{limits-u-c-100-10ab-1}, at least four orders of magnitude improvements have been obtained for $tqg$ couplings and at least one order of magnitude for all other FCNC couplings. In the light of the results on the limits on the FCNC coupling strengths in triple-top quark signal presented in Tables~\ref{limits-u-c-27-10-15-20ab-1} and \ref{limits-u-c-100-10ab-1}, we confirm the remarkable sensitivity that HE-LHC and FCC-hh would have on these couplings, especially on those of $tqg$ and $tqH$, which are comparatively much more constrained than other results in literature.

As we mentioned earlier, there are several sources of uncertainties that one needs to be taken into account to determine the 95\% CL upper limits. Following the strategy presented in Ref.~\cite{Liu:2020kxt}, we consider systematic uncertainty of 10\% for both the HE-LHC and FCC-hh. With this realistic 10\% systematic uncertainty, the sensitivities become slightly weaker than those without any systematic error which are presented in Table.~\ref{limits-u-c-27-10-15-20ab-1} and \ref{limits-u-c-100-10ab-1}. For the case of $tqg$ FCNC coupling at FCC-hh with  the integrated luminosity of 10 ab$^{-1}$, the sensitivity decreases to the $Br (t \to u g) = 5.43 \times 10^{-8}$ and $Br (t \to c g) = 9.87 \times 10^{-8}$. For the same couplings at  HE-LHC with the integrated luminosity of 15 ab$^{-1}$, the sensitivity decreases to the $Br (t \to u g) = 7.13 \times 10^{-8}$ and $Br (t \to c g) = 6.47 \times 10^{-8}$.

\begin{figure*}[htb]
\begin{center}
\vspace{0.40cm}
\resizebox{0.490\textwidth}{!}{\includegraphics{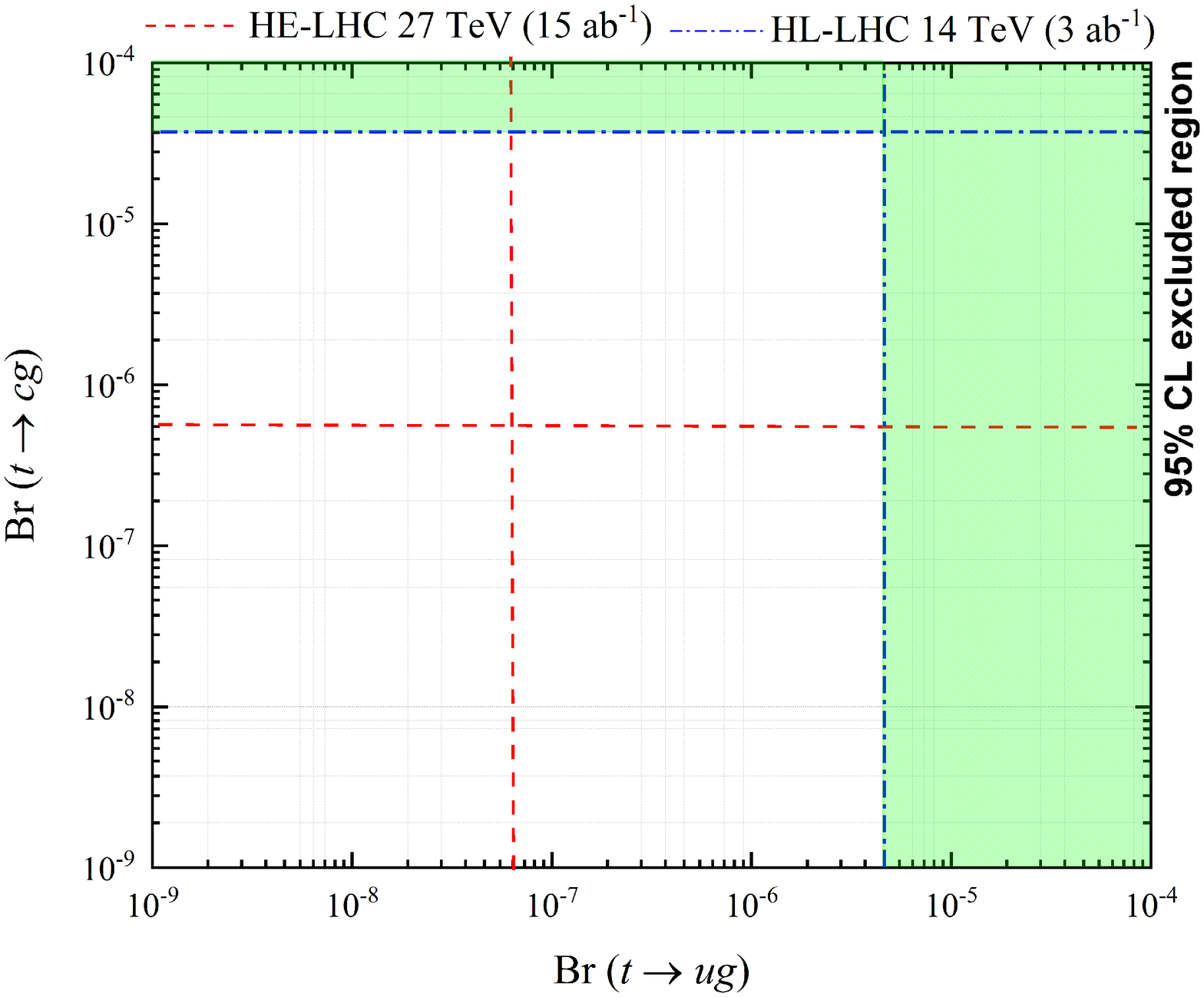}}   		
\resizebox{0.490\textwidth}{!}{\includegraphics{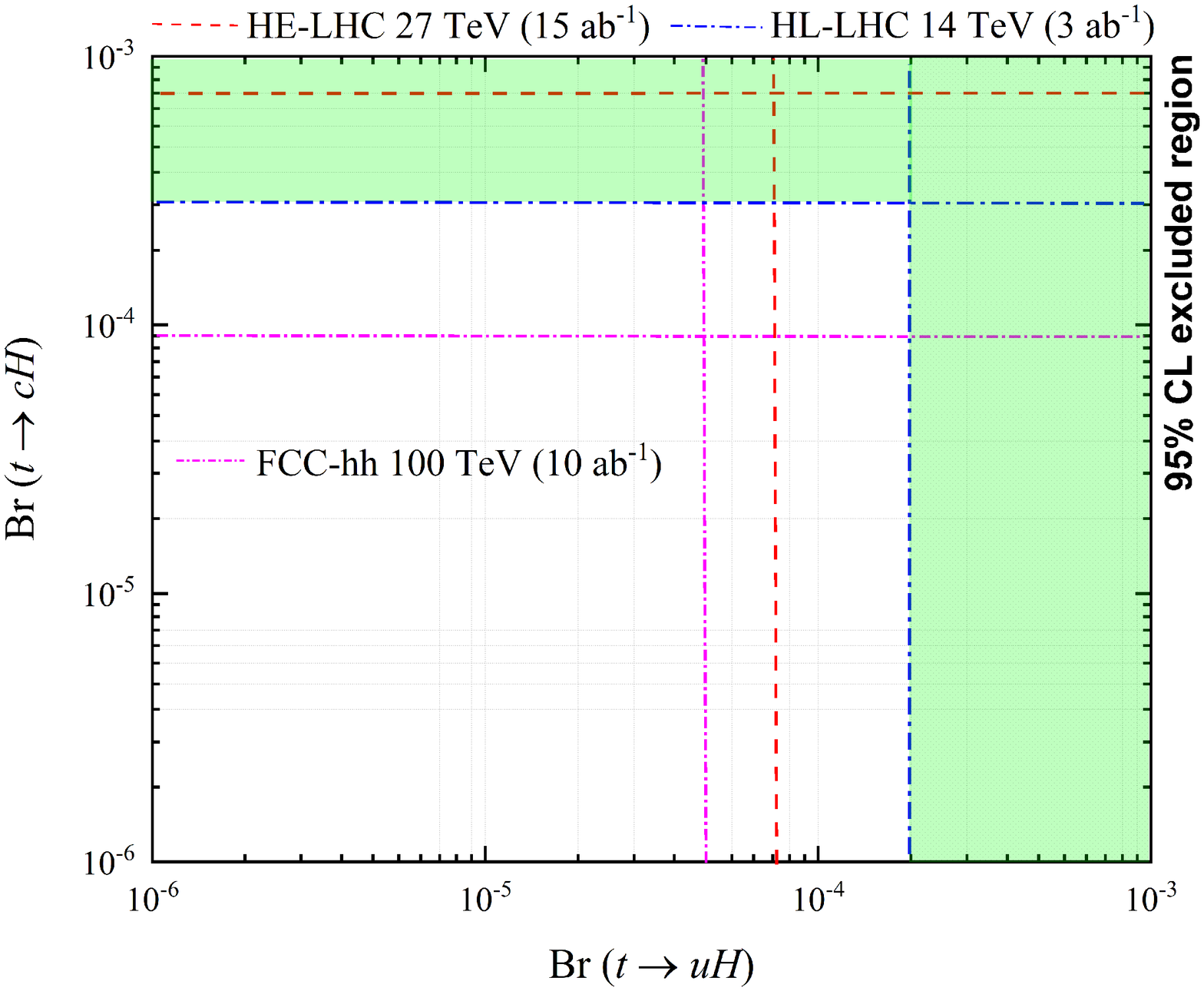}}   
\caption{ 
The 95\% CL expected exclusion limits on the $tqg$ and $tqH$ branching fractions for the HL-LHC with the upgraded CMS and ATLAS detector at an integrated luminosity of 3 ab$^{-1}$~\cite{CMS:2018kwi,ATLAS:HL-LHC}. The sensitivity of the 27 TeV HE-LHC with 15 ab$^{-1}$ and FCC-hh with 10 ab$^{-1}$ determined in this study are presented as well. } \label{fig:HE-LHC:HL-LHC}
\end{center}
\end{figure*}
\begin{figure}[htb]
\begin{center}
\vspace{0.40cm}
\resizebox{0.480\textwidth}{!}{\includegraphics{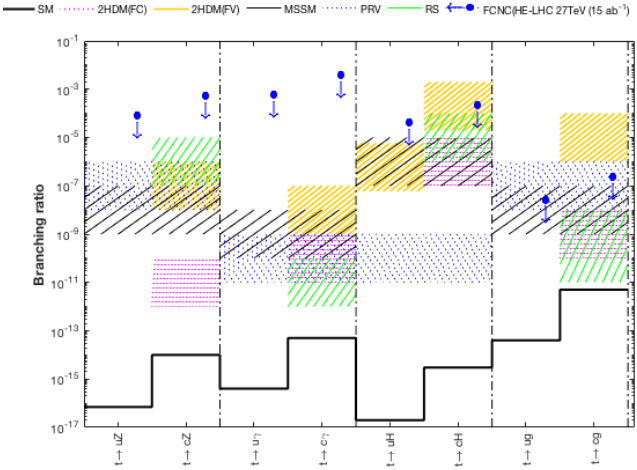}}   		
\caption{ Summary of branching fraction limits for top quark FCNC interactions obtained for the HE-LHC working at the integrated luminosity of 15 ab$^{-1}$, compared to the SM predictions, as well as to the various BSM predictions.  } \label{fig:Summary}
\end{center}
\end{figure}

To date, large amount of phenomenological studies available in literature have extensively investigated the association production of top quark with a gluon, photon, Higgs and $Z$ boson through different processes at future hadron and lepton colliders, see e.g. Refs.~\cite{Oyulmaz:2018irs,Denizli:2017cfx,Khanpour:2014xla,Hesari:2015oya,Oyulmaz:2019jqr,Behera:2018ryv,Aguilar-Saavedra:2017vka,Jain:2019ebq,Liu:2020kxt} for the most recent reviews.
It is worthwhile to compare the obtained limits in this study with those obtained by other groups. In Ref.~\cite{Liu:2020kxt}, the sensitivity of future searches for the top-Higgs boson FCNC couplings $tqh$ ($q=u, \, c$) at the HE-LHC and FCC-hh has been investigated. It is shown that the $Br(t \to qh)$ can be probed to  $Br(t \to uh)= 7.0 \times 10 ^{-5}$ and $Br(t \to ch) =  8.5 \times 10 ^{-5}$ at the HE-LHC with an integrated luminosity of  15 ab$^{-1}$, and $Br(t \to uh)= 2.3 \times 10 ^{-5}$ and $Br(t \to ch)= 3.0 \times 10 ^{-5}$ at the FCC-hh with an integrated luminosity of 30 ab$^{-1}$. In Ref.~\cite{Jain:2019ebq}, the prospects for discovering a top quark decaying into one light Higgs boson in top-quark pair production at the future hadron colliers have been presented. Based on their study, they found that with 27 TeV HE-LHC and 100 TeV FCC, the limits can reach up to $6.1 \times 10 ^{-4}$ and $2.0 \times 10 ^{-4}$, respectively. The top quark FCNC $tqg$ interactions at FCC-hh collider has been investigated in Ref.~\cite{Oyulmaz:2019jqr}. They found that the branching $Br(t \to qg)$ at the order of $10 ^{-7}$ can be achieved at FCC-hh with an integrated luminosity of 10 ab$^{-1}$.  In Ref.~\cite{Aguilar-Saavedra:2017vka}, searches for the anomalous $tqZ$ and $tq \gamma$ interactions have been done for the HE-LHC and FCC-hh. It has been shown that the expected limits on $t\to uZ/u\gamma$ branching ratios could reach the $10^{-5}$ level at HE-LHC and $10^{-6}$ level at FCC-hh.

As a final point, the present research explores for the first time, the sensitivity of triple-top signal at 27 TeV HE-LHC and 100 TeV FCC-hh to probe the top quark FCNC couplings with a gluon, photon, Higgs and $Z$ boson. The integrated luminosities from 10 to 20 ab$^{-1}$ are explored.  We have presented and discussed the sensitivity of such collider and shown that  the triple-top signal signature improves the current limits of LHC for most of top quark FCNC interactions, especially
for the $tqg$ and $tqH$ FCNC couplings. We have then examined and highlighted limits for HE-LHC working at different scenario of integrated luminosity. We have shown that for
the $tqg$ and $tqH$ modes, one can find a stronger sensitivity than that obtained by LHC and any future HL-LHC measurement by at least two to three order of magnitude.
In this study, we have introduced some methodological improvements aimed to improve the current limits on the top quark FCNC branching fractions. Regarding the results presented here, a number of important updates and improvements are foreseen. The main scope of the present paper is the beyond SM scenarios study of triple-top quark productions leading to anomalous $tqX$ vertices. Along with the phenomenological method presented here, additional observables can be use to suppress the background contributions and enhance the signal significance in extracting the reach of the couplings. 
A further improvement for future investigation is the use of multivariate technique (MVA)~\cite{Hocker:2007ht}. Such technique is expected to improve the limits of coupling strengths. Finally, one could also consider other channel of top quark production to study top quark FCNC transitions at the HE-LHC and FCC-hh. Taking into account the next to leading order (NLO) QCD corrections to the three top signal processes would tighten the upper bounds on the FCNC branching fractions. We leave investigations of such improvements to the future work.


%
\section{Summary and conclusions} \label{sec:Discussion}
%

We now turn to present our summary and conclusions. We first compare our limits with the projections of HL-LHC. Then, we conclude this section with a plot showing a summary of branching fraction limits for top quark FCNC interactions in compassion to the SM predictions, as well as to the various beyond SM scenarios.

All the proposed future lepton~\cite{Abada:2019zxq} and hadron~\cite{Benedikt:2018csr} colliders including the high energy large hadron collider (HE-LHC)~\cite{Abada:2019ono} will likely face critical decision making in the coming years~\cite{Abada:2019lih}. On the theory side, investigations of the top quark FCNC coupling to photons, gluon, Z and Higgs boson offer one of the important alternatives to explore new physics beyond the SM. Determination of new physics potentials of such proposed colliders at the energy frontier, in particular the reach on the top quark FCNC couplings measurements, is an important topic for the high energy physics community. Continued efforts are needed to be done to investigate the new physics accessible at these proposed colliders~\cite{FCC}. In recent years, there has been a considerable amount of literature that published to highlight the need for future high energy colliders.

Currently, the most stringent limits on top quark FCNC branching fractions $Br (t \to qX)$ have been measured with ATLAS and CMS Collaboration at the LHC. The results obtained at $\sqrt{s} = 13 \, {\text {TeV}}$ of LHC significantly improve the upper limits set with the 7 and 8 TeV data. It is worth mentioning here that, due to availability of large number of experimental results on the top quark FCNC transitions from the LHC through $pp$ collision, it could leads to a good prospects for pushing the top quark FCNC boundaries to even much higher constraints using future colliders. 

This paper focused to present phenomenological investigations to analyze the sensitivity of the triple-top quark signal at HE-LHC and FCC-hh to the top quark FCNC couplings, $tqX$ with $X=g, \, H, \, Z, \, \gamma$. To this end, we have studied the triple-top production $p p \to t \bar t t \, (\bar t t \bar t)$ at 27 TeV of HE-LHC and 100 TeV of FCC-hh taken into account the unique signal signature of two same-sign isolated charged lepton.

Numerical values are provided in Table~\ref{limits-u-c-27-10-15-20ab-1} and \ref{limits-u-c-100-10ab-1}. Considering the 27 TeV HE-LHC projections, our results clearly show that with an integrated luminosity of 15 ab$^{-1}$, it is likely that while the LHC can be obtained limits on the top quark FCNC couplings of $tqg$ and $tqH$ up to a sensitivity of the order $10^{-5}$ and $10^{-3}$, the limits on these couplings can reach up to a sensitivity of the order $10^{-8}$ and $10^{-5}$, respectively. Consistent with the literature, this study confirmed that most stringent limits can be obtained for the case of $tqg$.

The reach we obtain on the upper limits of top quark FCNC branching ratios $Br (t \to q g)$ and $Br (t \to q H)$ are highlighted in Fig.~\ref{fig:HE-LHC:HL-LHC}, for the two future hadron-hadron colliders considered in this analysis, HE-LHC and FCC-hh. 

As one can see from Fig.~\ref{fig:HE-LHC:HL-LHC}, our limit on the $tqg$ and $tqH$ branching ratios are much better than the projected limits on top
FCNC couplings at HL-LHC~\cite{CMS:2018kwi,ATLAS:HL-LHC}. While some of other calculated limits sound approximately similar to that are projected in the
case of HL-LHC, note that our analysis considered the triple-top signal which could let us to examine all the FCNC couplings. We believe that our study and the results presented here have clearly brought out the advantages of the HE-LHC and FCC-hh colliders in probing the top quark FCNC couplings with gluon, photon, $Z$ and Higgs boson, which could complement the information that could be extracted from the LHC and HL-LHC.

In order to have a better estimate of the extracted limits, let us conclude our discussions by comparing the branching fractions obtained in this study with theoretical predictions in the SM and other new physics models beyond SM. In Fig.~\ref{fig:Summary}, our obtained results are compared with the theoretical predictions in the SM, as well as with various new physics models. Concerning the limits obtained for the $t \to q g$ FCNC couplings, we observe that the limits are comparable or even better than the beyond SM predictions~\cite{Chiang:2018oyd}. The comparisons that shown for the obtained limits of the $tuH$ and $tcH$ deserve a separate comment. As can be seen in Fig.~\ref{fig:Summary}, the branching ratio for the $tqH$ couplings in the context of SM at one-loop level is of the order of $\sim$10$^{-15}$ which is far from the reach of current sensitivity of the LHC and future colliders. However, several models beyond SM predict the existence of FCNC transitions at the tree level~\cite{Chiang:2018oyd,Botella:2015hoa,Arroyo:2013tna,Celis:2015ara,Abbas:2015cua,Arroyo-Urena:2019qhl}. The predicted branching ratios of such models is ${\cal O} (10^{-3})$ as shown in Fig.~\ref{fig:Summary}. Hence, this opens the possibility that the experiments which will be carried out at the HE-LHC or FCC-hh can be done with high expectation for a detection. In comparison with those of new physics models, our findings for future projections of HE-LHC and FCC-hh in triple-top signal represent good prospects for pushing top FCNC boundaries to even higher constraints.

%
\section*{Acknowledgments}
%

We acknowledge fruitful discussions with Daniel Schulte, Mogens Dam,  Marco Zaro, Giovanni Zevi and Frederic Deliot in customizing the detector card used in this analysis. Author are thankful to Mojtaba Mohammadi Najafabadi for reading this manuscript and many helpful discussions and comments. We also gratefully acknowledge  Reza Goldouzian, Negin Shafie, Morteza Khatiri and Reza Jafari for useful comments and technical issues.  Author thanks School of Particles and Accelerators, Institute for Research in Fundamental Sciences (IPM) and University of Science and Technology of Mazandaran for financial support of this research. Author also is thankful the CERN theory department for their hospitality and support during the preparation of this paper.



%

\begin{thebibliography}{}
%




\bibitem{Tanabashi:2018oca} 
M.~Tanabashi {\it et al.} [Particle Data Group],
``Review of Particle Physics,''
Phys.\ Rev.\ D {\bf 98}, no. 3, 030001 (2018).
doi:10.1103/PhysRevD.98.030001







\bibitem{Husemann:2017eka} 
U.~Husemann,
``Top-Quark Physics: Status and Prospects,''
Prog.\ Part.\ Nucl.\ Phys.\  {\bf 95}, 48 (2017)
doi:10.1016/j.ppnp.2017.03.002
[arXiv:1704.01356 [hep-ex]].










\bibitem{Giammanco:2017xyn} 
A.~Giammanco and R.~Schwienhorst,
``Single top-quark production at the Tevatron and the LHC,''
Rev.\ Mod.\ Phys.\  {\bf 90}, no. 3, 035001 (2018)
doi:10.1103/RevModPhys.90.035001
[arXiv:1710.10699 [hep-ex]].











\bibitem{Cortiana:2015rca} 
G.~Cortiana,
``Top-quark mass measurements: review and perspectives,''
Rev.\ Phys.\  {\bf 1}, 60 (2016)
doi:10.1016/j.revip.2016.04.001
[arXiv:1510.04483 [hep-ex]].










\bibitem{Boos:2015bta} 
E.~Boos, O.~Brandt, D.~Denisov, S.~Denisov and P.~Grannis,
``The top quark (20 years after its discovery),''
Phys.\ Usp.\  {\bf 58}, no. 12, 1133 (2015)
[Usp.\ Fiz.\ Nauk {\bf 185}, no. 12, 1241 (2015)]
doi:10.3367/UFNe.0185.201512a.1241
[arXiv:1509.03325 [hep-ex]].










\bibitem{Cristinziani:2016vif} 
M.~Cristinziani and M.~Mulders,
``Top-quark physics at the Large Hadron Collider,''
J.\ Phys.\ G {\bf 44}, no. 6, 063001 (2017)
doi:10.1088/1361-6471/44/6/063001
[arXiv:1606.00327 [hep-ex]].











\bibitem{Englert:2017dev} 
C.~Englert and M.~Russell,
``Top quark electroweak couplings at future lepton colliders,''
Eur.\ Phys.\ J.\ C {\bf 77}, no. 8, 535 (2017)
doi:10.1140/epjc/s10052-017-5095-z
[arXiv:1704.01782 [hep-ph]].









\bibitem{Giammanco:2015bxk} 
A.~Giammanco,
``Single top quark production at the LHC,''
Rev.\ Phys.\  {\bf 1}, 1 (2016)
doi:10.1016/j.revip.2015.12.001
[arXiv:1511.06748 [hep-ex]].












\bibitem{Gouz:1998rk} 
Y.~P.~Gouz and S.~R.~Slabospitsky,
``Double top production at hadronic colliders,''
Phys.\ Lett.\ B {\bf 457}, 177 (1999)
doi:10.1016/S0370-2693(99)00516-X
[hep-ph/9811330].












\bibitem{Wicke:2010xw} 
D.~Wicke [CDF and D0 Collaborations],
``Single and Double Top Quark Production at the Tevatron,''
Nuovo Cim.\ C {\bf 033N5}, 263 (2010)
doi:10.1393/ncc/i2011-10709-1
[arXiv:1006.1275 [hep-ex]].












\bibitem{Glashow:1970gm} 
S.~L.~Glashow, J.~Iliopoulos and L.~Maiani,
``Weak Interactions with Lepton-Hadron Symmetry,''
Phys.\ Rev.\ D {\bf 2}, 1285 (1970).
doi:10.1103/PhysRevD.2.1285









\bibitem{AguilarSaavedra:2004wm} 
J.~A.~Aguilar-Saavedra,
``Top flavor-changing neutral interactions: Theoretical expectations and experimental detection,''
Acta Phys.\ Polon.\ B {\bf 35}, 2695 (2004)
[hep-ph/0409342].









\bibitem{Gaitan:2017tka} 
R.~Gaitán, R.~Martinez, J.~H.~M.~de Oca and E.~A.~Garcés,
``SM Higgs boson and $t\rightarrow cZ$ decays in the 2HDM type III with CP violation,''
Phys.\ Rev.\ D {\bf 98}, no. 3, 035031 (2018)
doi:10.1103/PhysRevD.98.035031
[arXiv:1710.04262 [hep-ph]].












\bibitem{Chiang:2018oyd} 
C.~W.~Chiang, U.~K.~Dey and T.~Jha,
``$t \rightarrow cg$ and $t \rightarrow cZ$ in universal extra-dimensional models,''
Eur.\ Phys.\ J.\ Plus {\bf 134}, no. 5, 210 (2019)
doi:10.1140/epjp/i2019-12607-1
[arXiv:1807.01481 [hep-ph]].












\bibitem{AguilarSaavedra:2002ns} 
J.~A.~Aguilar-Saavedra and B.~M.~Nobre,
``Rare top decays t ---> c gamma, t ---> cg and CKM unitarity,''
Phys.\ Lett.\ B {\bf 553}, 251 (2003)
doi:10.1016/S0370-2693(02)03230-6
[hep-ph/0210360].













\bibitem{Cao:2007dk} 
J.~J.~Cao, G.~Eilam, M.~Frank, K.~Hikasa, G.~L.~Liu, I.~Turan and J.~M.~Yang,
``SUSY-induced FCNC top-quark processes at the large hadron collider,''
Phys.\ Rev.\ D {\bf 75}, 075021 (2007)
doi:10.1103/PhysRevD.75.075021
[hep-ph/0702264].









\bibitem{Baum:2008qm} 
I.~Baum, G.~Eilam and S.~Bar-Shalom,
``Scalar flavor changing neutral currents and rare top quark decays in a two H iggs doublet model 'for the top quark',''
Phys.\ Rev.\ D {\bf 77}, 113008 (2008)
doi:10.1103/PhysRevD.77.113008
[arXiv:0802.2622 [hep-ph]].











\bibitem{Agashe:2004cp} 
K.~Agashe, G.~Perez and A.~Soni,
``Flavor structure of warped extra dimension models,''
Phys.\ Rev.\ D {\bf 71}, 016002 (2005)
doi:10.1103/PhysRevD.71.016002
[hep-ph/0408134].








\bibitem{Cao:2007bx} 
J.~j.~Cao, G.~l.~Liu, J.~M.~Yang and H.~j.~Zhang,
``Top-quark FCNC Productions at CERN LHC in Topcolor-assisted Technicolor Model,''
Phys.\ Rev.\ D {\bf 76}, 014004 (2007)
doi:10.1103/PhysRevD.76.014004
[hep-ph/0703308 [HEP-PH]].












\bibitem{Zhang:2007ub} 
H.~J.~Zhang,
``Top-quark FCNC decay t ---> cgg in topcolor-assisted technicolor model,''
Phys.\ Rev.\ D {\bf 77}, 057501 (2008)
doi:10.1103/PhysRevD.77.057501
[arXiv:0712.0151 [hep-ph]].













\bibitem{Oyulmaz:2018irs} 
K.~Y.~Oyulmaz, A.~Senol, H.~Denizli, A.~Yilmaz, I.~Turk Cakir and O.~Cakir,
``Probing anomalous $tq\gamma$ and $tqg$ couplings via single top production in association with photon at FCC-hh,''
Eur.\ Phys.\ J.\ C {\bf 79}, no. 1, 83 (2019)
doi:10.1140/epjc/s10052-019-6593-y
[arXiv:1811.01074 [hep-ph]].











\bibitem{Khatibi:2014via} 
S.~Khatibi and M.~Mohammadi Najafabadi,
``Probing the Anomalous FCNC Interactions in Top-Higgs Final State and Charge Ratio Approach,''
Phys.\ Rev.\ D {\bf 89}, no. 5, 054011 (2014)
doi:10.1103/PhysRevD.89.054011
[arXiv:1402.3073 [hep-ph]].












\bibitem{Khatibi:2014bsa} 
S.~Khatibi and M.~Mohammadi Najafabadi,
``Exploring the Anomalous Higgs-top Couplings,''
Phys.\ Rev.\ D {\bf 90}, no. 7, 074014 (2014)
doi:10.1103/PhysRevD.90.074014
[arXiv:1409.6553 [hep-ph]].









\bibitem{Khatibi:2015aal} 
S.~Khatibi and M.~Mohammadi Najafabadi,
``Constraints on top quark flavor changing neutral currents using diphoton events at the LHC,''
Nucl.\ Phys.\ B {\bf 909}, 607 (2016)
doi:10.1016/j.nuclphysb.2016.06.009
[arXiv:1511.00220 [hep-ph]].







\bibitem{TurkCakir:2017rvu} 
I.~Turk Cakir, A.~Yilmaz, H.~Denizli, A.~Senol, H.~Karadeniz and O.~Cakir,
``Probing the Anomalous FCNC Couplings at Large Hadron Electron Collider,''
Adv.\ High Energy Phys.\  {\bf 2017}, 1572053 (2017)
doi:10.1155/2017/1572053
[arXiv:1705.05419 [hep-ph]].











\bibitem{Denizli:2017cfx} 
H.~Denizli, A.~Senol, A.~Yilmaz, I.~Turk Cakir, H.~Karadeniz and O.~Cakir,
``Top quark FCNC couplings at future circular hadron electron colliders,''
Phys.\ Rev.\ D {\bf 96}, no. 1, 015024 (2017)
doi:10.1103/PhysRevD.96.015024
[arXiv:1701.06932 [hep-ph]].










\bibitem{Khanpour:2014xla} 
H.~Khanpour, S.~Khatibi, M.~Khatiri Yanehsari and M.~Mohammadi Najafabadi,
``Single top quark production as a probe of anomalous $tq\gamma$ and $tqZ$ couplings at the FCC-ee,''
Phys.\ Lett.\ B {\bf 775}, 25 (2017)
doi:10.1016/j.physletb.2017.10.047
[arXiv:1408.2090 [hep-ph]].










\bibitem{Hesari:2015oya} 
H.~Hesari, H.~Khanpour and M.~Mohammadi Najafabadi,
``Direct and Indirect Searches for Top-Higgs FCNC Couplings,''
Phys.\ Rev.\ D {\bf 92}, no. 11, 113012 (2015)
doi:10.1103/PhysRevD.92.113012
[arXiv:1508.07579 [hep-ph]].









\bibitem{Shi:2019epw} 
L.~Shi and C.~Zhang,
``Probing the top quark flavor-changing couplings at CEPC,''
Chin.\ Phys.\ C {\bf 43}, no. 11, 113104 (2019)
doi:10.1088/1674-1137/43/11/113104
[arXiv:1906.04573 [hep-ph]].










\bibitem{Alici:2019asv} 
E.~Alici and M.~Köksal,
``Probing the anomalous $tq\gamma$ couplings through single top production at the future lepton-hadron colliders,''
Mod.\ Phys.\ Lett.\ A {\bf 34}, no. 36, 1950298 (2019)
doi:10.1142/S0217732319502985
[arXiv:1905.00588 [hep-ph]].








\bibitem{Oyulmaz:2019jqr} 
K.~Y.~Oyulmaz, A.~Senol, H.~Denizli and O.~Cakir,
``Top quark anomalous FCNC production via $tqg$ couplings at FCC-hh,''
Phys.\ Rev.\ D {\bf 99}, no. 11, 115023 (2019)
doi:10.1103/PhysRevD.99.115023
[arXiv:1902.03037 [hep-ph]].









\bibitem{Behera:2018ryv} 
S.~Behera, R.~Islam, M.~Kumar, P.~Poulose and R.~Rahaman,
``Fingerprinting the Top quark FCNC via anomalous $Ztq$ couplings at the LHeC,''
Phys.\ Rev.\ D {\bf 100}, no. 1, 015006 (2019)
doi:10.1103/PhysRevD.100.015006
[arXiv:1811.04681 [hep-ph]].







\bibitem{Aguilar-Saavedra:2017vka} 
J.~A.~Aguilar-Saavedra,
``Ultraboosted $Zt$ and $\gamma t$ production at the HL-LHC and FCC-hh,''
Eur.\ Phys.\ J.\ C {\bf 77}, no. 11, 769 (2017)
doi:10.1140/epjc/s10052-017-5375-7
[arXiv:1709.03975 [hep-ph]].











\bibitem{Shen:2018mlj} 
J.~F.~Shen, Y.~Q.~Li and Y.~B.~Liu,
``Searches for anomalous tqZ couplings from the trilepton signal of tZ associated production at the 14 TeV LHC,''
Phys.\ Lett.\ B {\bf 776}, 391 (2018)
doi:10.1016/j.physletb.2017.11.055
[arXiv:1712.03506 [hep-ph]].









\bibitem{AguilarSaavedra:2010rx} 
J.~A.~Aguilar-Saavedra,
``Zt, gamma t and t production at hadron colliders via strong flavour-changing neutral couplings,''
Nucl.\ Phys.\ B {\bf 837}, 122 (2010)
doi:10.1016/j.nuclphysb.2010.05.005
[arXiv:1003.3173 [hep-ph]].






\bibitem{Jain:2019ebq} 
R.~Jain and C.~Kao,
``Charming top decays with a flavor changing neutral Higgs boson and $WW$ at hadron colliders,''
Phys.\ Rev.\ D {\bf 99}, no. 5, 055036 (2019),
[arXiv:1901.00157 [hep-ph]].






\bibitem{Liu:2020kxt} 
Y.~B.~Liu and S.~Moretti,
``Probing the top-Higgs boson FCNC couplings via the $h\to \gamma\gamma$ channel at the HE-LHC and FCC-hh,''
Phys.\ Rev.\ D {\bf 101}, no. 7, 075029 (2020),
[arXiv:2002.05311 [hep-ph]].










\bibitem{CMS:1900mtx} 
CMS Collaboration [CMS Collaboration],
``Projections of sensitivities for tttt production at HL-LHC and HE-LHC,''
CMS-PAS-FTR-18-031.







\bibitem{Cao:2019qrb} 
Q.~H.~Cao, S.~L.~Chen, Y.~Liu and X.~P.~Wang,
``What can We Learn from Triple Top-Quark Production?,''
Phys.\ Rev.\ D {\bf 100}, no. 5, 055035 (2019)
doi:10.1103/PhysRevD.100.055035
[arXiv:1901.04643 [hep-ph]].









\bibitem{CidVidal:2018eel} 
X.~Cid Vidal {\it et al.},
``Report from Working Group 3 : Beyond the Standard Model physics at the HL-LHC and HE-LHC,''
CERN Yellow Rep.\ Monogr.\  {\bf 7}, 585 (2019)
doi:10.23731/CYRM-2019-007.585
[arXiv:1812.07831 [hep-ph]].









\bibitem{Cepeda:2019klc} 
M.~Cepeda {\it et al.},
``Report from Working Group 2 : Higgs Physics at the HL-LHC and HE-LHC,''
CERN Yellow Rep.\ Monogr.\  {\bf 7}, 221 (2019)
doi:10.23731/CYRM-2019-007.221
[arXiv:1902.00134 [hep-ph]].









\bibitem{Azzi:2019yne} 
P.~Azzi {\it et al.},
``Report from Working Group 1 : Standard Model Physics at the HL-LHC and HE-LHC,''
CERN Yellow Rep.\ Monogr.\  {\bf 7}, 1 (2019)
doi:10.23731/CYRM-2019-007.1
[arXiv:1902.04070 [hep-ph]].










\bibitem{CMS:2018kwi} 
CMS Collaboration [CMS Collaboration],
``Prospects for the search for gluon-mediated FCNC in top quark production with the CMS Phase-2 detector at the HL-LHC,''
CMS-PAS-FTR-18-004.







\bibitem{ATLAS:HL-LHC} 
Expected sensitivity of ATLAS to FCNC top quark decays $t \to Zq$ and $t \to Hq$ at the High Luminosity LHC, https://cds.cern.ch/record/2209126, ATL-PHYS-PUB-2016-019







\bibitem{Cerri:2018ypt} 
A.~Cerri {\it et al.},
``Report from Working Group 4 : Opportunities in Flavour Physics at the HL-LHC and HE-LHC,''
CERN Yellow Rep.\ Monogr.\  {\bf 7}, 867 (2019)
doi:10.23731/CYRM-2019-007.867
[arXiv:1812.07638 [hep-ph]].












\bibitem{Benedikt:2018csr} 
A.~Abada {\it et al.} [FCC Collaboration],
``FCC-hh: The Hadron Collider : Future Circular Collider Conceptual Design Report Volume 3,''
Eur.\ Phys.\ J.\ ST {\bf 228}, no. 4, 755 (2019).
doi:10.1140/epjst/e2019-900087-0









\bibitem{Mangano:2017tke} 
M.~Mangano,
``Physics at the FCC-hh, a 100 TeV pp collider,''
doi:10.23731/CYRM-2017-003
arXiv:1710.06353 [hep-ph].










\bibitem{Arkani-Hamed:2015vfh} 
N.~Arkani-Hamed, T.~Han, M.~Mangano and L.~T.~Wang,
``Physics opportunities of a 100 TeV proton?proton collider,''
Phys.\ Rept.\  {\bf 652}, 1 (2016)
doi:10.1016/j.physrep.2016.07.004
[arXiv:1511.06495 [hep-ph]].










\bibitem{Aad:2015gea} 
G.~Aad {\it et al.} [ATLAS Collaboration],
``Search for single top-quark production via flavour-changing neutral currents at 8 TeV with the ATLAS detector,''
Eur.\ Phys.\ J.\ C {\bf 76}, no. 2, 55 (2016)
doi:10.1140/epjc/s10052-016-3876-4
[arXiv:1509.00294 [hep-ex]].









\bibitem{Aaboud:2018pob} 
M.~Aaboud {\it et al.} [ATLAS Collaboration],
``Search for flavor-changing neutral currents in top quark decays $t\to Hc$ and $t \to Hu$ in multilepton final states in proton-proton collisions at $\sqrt{s}= 13$ TeV with the ATLAS detector,''
Phys.\ Rev.\ D {\bf 98}, no. 3, 032002 (2018)
doi:10.1103/PhysRevD.98.032002
[arXiv:1805.03483 [hep-ex]].











\bibitem{CMS:2017twu} 
CMS Collaboration [CMS Collaboration],
``Search for flavour changing neutral currents in top quark production and decays with three-lepton final state using the data collected at sqrt(s) = 13 TeV,''
CMS-PAS-TOP-17-017.









\bibitem{Khachatryan:2015att} 
V.~Khachatryan {\it et al.} [CMS Collaboration],
``Search for Anomalous Single Top Quark Production in Association with a Photon in $pp$ Collisions at $ \sqrt{s}=8 $ TeV,''
JHEP {\bf 1604}, 035 (2016)
doi:10.1007/JHEP04(2016)035
[arXiv:1511.03951 [hep-ex]].










\bibitem{Sirunyan:2017kkr} 
A.~M.~Sirunyan {\it et al.} [CMS Collaboration],
``Search for associated production of a Z boson with a single top quark and for tZ flavour-changing interactions in pp collisions at $ \sqrt{s}=8 $ TeV,''
JHEP {\bf 1707}, 003 (2017)
doi:10.1007/JHEP07(2017)003
[arXiv:1702.01404 [hep-ex]].








\bibitem{Aaboud:2017mfd} 
M.~Aaboud {\it et al.} [ATLAS Collaboration],
``Search for top quark decays $t\rightarrow qH$, with $H\to\gamma\gamma$, in $\sqrt{s}=13$ TeV $pp$ collisions using the ATLAS detector,''
JHEP {\bf 1710}, 129 (2017)
doi:10.1007/JHEP10(2017)129
[arXiv:1707.01404 [hep-ex]].












\bibitem{Khachatryan:2016sib} 
V.~Khachatryan {\it et al.} [CMS Collaboration],
``Search for anomalous Wtb couplings and flavour-changing neutral currents in t-channel single top quark production in pp collisions at $\sqrt{s} =$ 7 and 8 TeV,''
JHEP {\bf 1702}, 028 (2017)
doi:10.1007/JHEP02(2017)028
[arXiv:1610.03545 [hep-ex]].









\bibitem{CMS:2016bss} 
CMS Collaboration [CMS Collaboration],
``Search for associated production of a Z boson with a single top quark and for tZ flavour-changing interactions in pp collisions at sqrt(s) = 8 TeV,''
CMS-PAS-TOP-12-039.










\bibitem{Aaboud:2018nyl} 
M.~Aaboud {\it et al.} [ATLAS Collaboration],
``Search for flavour-changing neutral current top-quark decays $t\to qZ$ in proton-proton collisions at $\sqrt{s}=13$ TeV with the ATLAS detector,''
JHEP {\bf 1807}, 176 (2018)
doi:10.1007/JHEP07(2018)176
[arXiv:1803.09923 [hep-ex]].









\bibitem{Sirunyan:2017nbr} 
A.~M.~Sirunyan {\it et al.} [CMS Collaboration],
``Measurement of the associated production of a single top quark and a Z boson in pp collisions at $\sqrt{s} =$ 13 TeV,''
Phys.\ Lett.\ B {\bf 779}, 358 (2018)
doi:10.1016/j.physletb.2018.02.025
[arXiv:1712.02825 [hep-ex]].








\bibitem{Aaboud:2017ylb} 
M.~Aaboud {\it et al.} [ATLAS Collaboration],
``Measurement of the production cross-section of a single top quark in association with a Z boson in proton?proton collisions at 13 TeV with the ATLAS detector,''
Phys.\ Lett.\ B {\bf 780}, 557 (2018)
doi:10.1016/j.physletb.2018.03.023
[arXiv:1710.03659 [hep-ex]].







\bibitem{Aaboud:2018oqm} 
M.~Aaboud {\it et al.} [ATLAS Collaboration],
``Search for top-quark decays $t \to Hq$ with 36 fb$^{-1}$ of $pp$ collision data at $\sqrt{s}=13$ TeV with the ATLAS detector,''
JHEP {\bf 1905}, 123 (2019)
doi:10.1007/JHEP05(2019)123
[arXiv:1812.11568 [hep-ex]].








\bibitem{Sirunyan:2017uae} 
A.~M.~Sirunyan {\it et al.} [CMS Collaboration],
``Search for the flavor-changing neutral current interactions of the top quark and the Higgs boson which decays into a pair of b quarks at $\sqrt{s}=$ 13 TeV,''
JHEP {\bf 1806}, 102 (2018)
doi:10.1007/JHEP06(2018)102
[arXiv:1712.02399 [hep-ex]].






\bibitem{Aad:2015pja} 
G.~Aad {\it et al.} [ATLAS Collaboration],
``Search for flavour-changing neutral current top quark decays $t\to Hq$ in $pp$ collisions at $\sqrt{s}=8$ TeV with the ATLAS detector,''
JHEP {\bf 1512}, 061 (2015)
doi:10.1007/JHEP12(2015)061
[arXiv:1509.06047 [hep-ex]].









\bibitem{AguilarSaavedra:2008zc} 
J.~A.~Aguilar-Saavedra,
``A Minimal set of top anomalous couplings,''
Nucl.\ Phys.\ B {\bf 812}, 181 (2009)
doi:10.1016/j.nuclphysb.2008.12.012
[arXiv:0811.3842 [hep-ph]].








\bibitem{Atlas:2019qfx} 
ATLAS and CMS Collaborations [ATLAS and CMS Collaborations],
``Addendum to the report on the physics at the HL-LHC, and perspectives for the HE-LHC: Collection of notes from ATLAS and CMS,''
CERN Yellow Rep.\ Monogr.\  {\bf 7}, Addendum (2019)
doi:10.23731/CYRM-2019-007.Addendum
[arXiv:1902.10229 [hep-ex]].








\bibitem{AguilarSaavedra:2009mx} 
J.~A.~Aguilar-Saavedra,
``A Minimal set of top-Higgs anomalous couplings,''
Nucl.\ Phys.\ B {\bf 821}, 215 (2009)
doi:10.1016/j.nuclphysb.2009.06.022
[arXiv:0904.2387 [hep-ph]].







\bibitem{Goldouzian:2014nha} 
R.~Goldouzian,
``Search for top quark flavor changing neutral currents in same-sign top quark production,''
Phys.\ Rev.\ D {\bf 91}, no. 1, 014022 (2015)
doi:10.1103/PhysRevD.91.014022
[arXiv:1408.0493 [hep-ph]].






\bibitem{Goldouzian:2016mrt} 
R.~Goldouzian and B.~Clerbaux,
``Photon initiated single top quark production via flavor-changing neutral currents at the LHC,''
Phys.\ Rev.\ D {\bf 95}, no. 5, 054014 (2017)
doi:10.1103/PhysRevD.95.054014
[arXiv:1609.04838 [hep-ph]].







\bibitem{Buschmann:2016uzg} 
M.~Buschmann, J.~Kopp, J.~Liu and X.~P.~Wang,
``New Signatures of Flavor Violating Higgs Couplings,''
JHEP {\bf 1606}, 149 (2016)
doi:10.1007/JHEP06(2016)149
[arXiv:1601.02616 [hep-ph]].











\bibitem{delAguila:1999kfp} 
F.~del Aguila and J.~A.~Aguilar-Saavedra,
``Multilepton production via top flavor changing neutral couplings at the CERN LHC,''
Nucl.\ Phys.\ B {\bf 576}, 56 (2000)
doi:10.1016/S0550-3213(00)00100-0
[hep-ph/9909222].







\bibitem{Sun:2016kek} 
H.~Sun and X.~Wang,
``Exploring the Anomalous Top-Higgs FCNC Couplings at the electron proton colliders,''
Eur.\ Phys.\ J.\ C {\bf 78}, no. 4, 281 (2018)
doi:10.1140/epjc/s10052-018-5761-9
[arXiv:1602.04670 [hep-ph]].








\bibitem{Guo:2016kea} 
Y.~C.~Guo, C.~X.~Yue and S.~Yang,
``Search for anomalous couplings via single top quark production in association with a photon at LHC,''
Eur.\ Phys.\ J.\ C {\bf 76}, no. 11, 596 (2016)
doi:10.1140/epjc/s10052-016-4452-7
[arXiv:1603.00604 [hep-ph]].








\bibitem{Liu:2016gsi} 
Y.~B.~Liu and Z.~J.~Xiao,
``Searches for the FCNC couplings from top-Higgs associated production signal with $h\to \gamma\gamma$ at the LHC,''
Phys.\ Lett.\ B {\bf 763}, 458 (2016)
doi:10.1016/j.physletb.2016.11.004
[arXiv:1610.03250 [hep-ph]].






\bibitem{Liu:2016dag} 
Y.~B.~Liu and Z.~J.~Xiao,
``Searches for top-Higgs FCNC couplings via the Whj signal with $h\to \gamma\gamma$ at the LHC,''
Phys.\ Rev.\ D {\bf 94}, no. 5, 054018 (2016)
doi:10.1103/PhysRevD.94.054018
[arXiv:1605.01179 [hep-ph]].







\bibitem{Liu:2019wmi} 
W.~Liu and H.~Sun,
``Top FCNC interactions through dimension six four-fermion operators at the electron proton collider,''
Phys.\ Rev.\ D {\bf 100}, no. 1, 015011 (2019)
doi:10.1103/PhysRevD.100.015011
[arXiv:1906.04884 [hep-ph]].








\bibitem{Kumbhakar:2019njm} 
S.~Kumbhakar and J.~Saini,
``Flavor signatures of complex anomalous $tcZ$ couplings,''
Eur.\ Phys.\ J.\ Plus {\bf 135}, no. 3, 330 (2020)
doi:10.1140/epjp/s13360-020-00341-8
[arXiv:1905.07690 [hep-ph]].











\bibitem{Malekhosseini:2018fgp} 
M.~Malekhosseini, M.~Ghominejad, H.~Khanpour and M.~Mohammadi Najafabadi,
``Constraining top quark flavor violation and dipole moments through three and four-top quark productions at the LHC,''
Phys.\ Rev.\ D {\bf 98}, no. 9, 095001 (2018)
doi:10.1103/PhysRevD.98.095001
[arXiv:1804.05598 [hep-ph]].










\bibitem{Papaefstathiou:2017xuv} 
A.~Papaefstathiou and G.~Tetlalmatzi-Xolocotzi,
``Rare top quark decays at a 100 TeV proton?proton collider: $t \rightarrow bWZ$ and $t\rightarrow hc$,''
Eur.\ Phys.\ J.\ C {\bf 78}, no. 3, 214 (2018)
doi:10.1140/epjc/s10052-018-5701-8
[arXiv:1712.06332 [hep-ph]].








\bibitem{Dey:2016cve} 
U.~K.~Dey and T.~Jha,
``Rare top decays in minimal and nonminimal universal extra dimension models,''
Phys.\ Rev.\ D {\bf 94}, no. 5, 056011 (2016)
doi:10.1103/PhysRevD.94.056011
[arXiv:1602.03286 [hep-ph]].









\bibitem{Barger:2010uw} 
V.~Barger, W.~Y.~Keung and B.~Yencho,
``Triple-Top Signal of New Physics at the LHC,''
Phys.\ Lett.\ B {\bf 687}, 70 (2010)
doi:10.1016/j.physletb.2010.03.001
[arXiv:1001.0221 [hep-ph]].








\bibitem{Chen:2014ewl} 
C.~R.~Chen,
``Searching for new physics with triple-top signal at the LHC,''
Phys.\ Lett.\ B {\bf 736}, 321 (2014).
doi:10.1016/j.physletb.2014.07.041










\bibitem{Chatrchyan:2012fla} 
S.~Chatrchyan {\it et al.} [CMS Collaboration],
``Search for heavy Majorana Neutrinos in $\mu^{\pm}\mu^{\pm} +$ Jets and $e^{\pm}e^{\pm} +$ Jets Events in pp Collisions at $\sqrt{s} =$ 7 TeV,''
Phys.\ Lett.\ B {\bf 717}, 109 (2012)
doi:10.1016/j.physletb.2012.09.012
[arXiv:1207.6079 [hep-ex]].








\bibitem{Alloul:2013bka} 
A.~Alloul, N.~D.~Christensen, C.~Degrande, C.~Duhr and B.~Fuks,
``FeynRules  2.0 - A complete toolbox for tree-level phenomenology,''
Comput.\ Phys.\ Commun.\  {\bf 185}, 2250 (2014)
doi:10.1016/j.cpc.2014.04.012
[arXiv:1310.1921 [hep-ph]].








\bibitem{Degrande:2011ua} 
C.~Degrande, C.~Duhr, B.~Fuks, D.~Grellscheid, O.~Mattelaer and T.~Reiter,
``UFO - The Universal FeynRules Output,''
Comput.\ Phys.\ Commun.\  {\bf 183}, 1201 (2012)
doi:10.1016/j.cpc.2012.01.022
[arXiv:1108.2040 [hep-ph]].







\bibitem{Alwall:2014hca} 
J.~Alwall {\it et al.},
``The automated computation of tree-level and next-to-leading order differential cross sections, and their matching to parton shower simulations,''
JHEP {\bf 1407}, 079 (2014)
doi:10.1007/JHEP07(2014)079
[arXiv:1405.0301 [hep-ph]].







\bibitem{Frederix:2018nkq} 
R.~Frederix, S.~Frixione, V.~Hirschi, D.~Pagani, H.-S.~Shao and M.~Zaro,
``The automation of next-to-leading order electroweak calculations,''
JHEP {\bf 1807}, 185 (2018)
doi:10.1007/JHEP07(2018)185
[arXiv:1804.10017 [hep-ph]].







\bibitem{Ball:2017nwa} 
R.~D.~Ball {\it et al.} [NNPDF Collaboration],
``Parton distributions from high-precision collider data,''
Eur.\ Phys.\ J.\ C {\bf 77}, no. 10, 663 (2017)
doi:10.1140/epjc/s10052-017-5199-5
[arXiv:1706.00428 [hep-ph]].









\bibitem{Ball:2014uwa} 
R.~D.~Ball {\it et al.} [NNPDF Collaboration],
``Parton distributions for the LHC Run II,''
JHEP {\bf 1504}, 040 (2015)
doi:10.1007/JHEP04(2015)040
[arXiv:1410.8849 [hep-ph]].










\bibitem{Ball:2012cx} 
R.~D.~Ball {\it et al.},
``Parton distributions with LHC data,''
Nucl.\ Phys.\ B {\bf 867}, 244 (2013)
doi:10.1016/j.nuclphysb.2012.10.003
[arXiv:1207.1303 [hep-ph]].








\bibitem{Sjostrand:2014zea} 
T.~Sjöstrand {\it et al.},
``An Introduction to PYTHIA 8.2,''
Comput.\ Phys.\ Commun.\  {\bf 191}, 159 (2015)
doi:10.1016/j.cpc.2015.01.024
[arXiv:1410.3012 [hep-ph]].









\bibitem{Cacciari:2011ma} 
M.~Cacciari, G.~P.~Salam and G.~Soyez,
``FastJet User Manual,''
Eur.\ Phys.\ J.\ C {\bf 72}, 1896 (2012)
doi:10.1140/epjc/s10052-012-1896-2
[arXiv:1111.6097 [hep-ph]].








\bibitem{Cacciari:2008gp} 
M.~Cacciari, G.~P.~Salam and G.~Soyez,
``The anti-$k_t$ jet clustering algorithm,''
JHEP {\bf 0804}, 063 (2008)
doi:10.1088/1126-6708/2008/04/063
[arXiv:0802.1189 [hep-ph]].








\bibitem{deFavereau:2013fsa} 
J.~de Favereau {\it et al.} [DELPHES 3 Collaboration],
``DELPHES 3, A modular framework for fast simulation of a generic collider experiment,''
JHEP {\bf 1402}, 057 (2014)
doi:10.1007/JHEP02(2014)057
[arXiv:1307.6346 [hep-ex]].














\bibitem{cla} 
\url{https://twiki.cern.ch/twiki/bin/view/CMS/RooStatsCl95}\,,
\url{https://twiki.cern.ch/twiki/bin/viewauth/CMS/StatisticsTools}







\bibitem{Sirunyan:2017uzs} 
A.~M.~Sirunyan {\it et al.} [CMS Collaboration],
``Measurement of the cross section for top quark pair production in association with a W or Z boson in proton-proton collisions at $\sqrt{s} =$ 13 TeV,''
JHEP {\bf 1808}, 011 (2018),
[arXiv:1711.02547 [hep-ex]].





\bibitem{CMS:2019too} 
A.~M.~Sirunyan {\it et al.} [CMS Collaboration],
``Measurement of top quark pair production in association with a Z boson in proton-proton collisions at $\sqrt{s}=$ 13 TeV,''
JHEP {\bf 2003}, 056 (2020),
[arXiv:1907.11270 [hep-ex]].







\bibitem{Hocker:2007ht} 
A.~Hocker {\it et al.},
``TMVA - Toolkit for Multivariate Data Analysis,''
physics/0703039 [physics.data-an].









\bibitem{Abada:2019zxq} 
A.~Abada {\it et al.} [FCC Collaboration],
``FCC-ee: The Lepton Collider : Future Circular Collider Conceptual Design Report Volume 2,''
Eur.\ Phys.\ J.\ ST {\bf 228}, no. 2, 261 (2019).
doi:10.1140/epjst/e2019-900045-4







\bibitem{Abada:2019ono} 
A.~Abada {\it et al.} [FCC Collaboration],
``HE-LHC: The High-Energy Large Hadron Collider : Future Circular Collider Conceptual Design Report Volume 4,''
Eur.\ Phys.\ J.\ ST {\bf 228}, no. 5, 1109 (2019).
doi:10.1140/epjst/e2019-900088-6









\bibitem{Abada:2019lih} 
A.~Abada {\it et al.} [FCC Collaboration],
``FCC Physics Opportunities : Future Circular Collider Conceptual Design Report Volume 1,''
Eur.\ Phys.\ J.\ C {\bf 79}, no. 6, 474 (2019).
doi:10.1140/epjc/s10052-019-6904-3








\bibitem{FCC} 
M. Benedikt, A. Blondel, P. Janot, M. Klein, M. Mangano, M. Mccullough, V. Mertens, K. Oide, W. Riegler, D. Schulte, and F. Zimmermann,
``Future Circular Colliders,''
Ann.\ Rev.\ Nucl.\ Part.\ Sci.\  {\bf 69} 2019






\bibitem{Botella:2015hoa} 
F.~J.~Botella, G.~C.~Branco, M.~Nebot and M.~N.~Rebelo,
``Flavour Changing Higgs Couplings in a Class of Two Higgs Doublet Models,''
Eur.\ Phys.\ J.\ C {\bf 76}, no. 3, 161 (2016)
doi:10.1140/epjc/s10052-016-3993-0
[arXiv:1508.05101 [hep-ph]].









\bibitem{Arroyo:2013tna} 
M.~A.~Arroyo-Ureña, J.~L.~Diaz-Cruz, E.~Díaz and J.~A.~Orduz-Ducuara,
``Flavor violating Higgs signals in the Texturized Two-Higgs Doublet Model (THDM-Tx),''
Chin.\ Phys.\ C {\bf 40}, no. 12, 123103 (2016)
doi:10.1088/1674-1137/40/12/123103
[arXiv:1306.2343 [hep-ph]].







\bibitem{Celis:2015ara} 
A.~Celis, J.~Fuentes-Martin, M.~Jung and H.~Serodio,
``Family nonuniversal Z? models with protected flavor-changing interactions,''
Phys.\ Rev.\ D {\bf 92}, no. 1, 015007 (2015)
doi:10.1103/PhysRevD.92.015007
[arXiv:1505.03079 [hep-ph]].









\bibitem{Abbas:2015cua} 
G.~Abbas, A.~Celis, X.~Q.~Li, J.~Lu and A.~Pich,
``Flavour-changing top decays in the aligned two-Higgs-doublet model,''
JHEP {\bf 1506}, 005 (2015)
doi:10.1007/JHEP06(2015)005
[arXiv:1503.06423 [hep-ph]].






\bibitem{Arroyo-Urena:2019qhl} 
M.~A.~Arroyo-Ureña, R.~Gaitán-Lozano, E.~A.~Herrera-Chacón, J.~H.~Montes de Oca Y. and T.~A.~Valencia-Pérez,
``Search for the $t\to ch$ decay at hadron colliders,''
JHEP {\bf 1907}, 041 (2019)
doi:10.1007/JHEP07(2019)041
[arXiv:1903.02718 [hep-ph]].




	
	
\end{thebibliography}
%


%

\end{document}